\documentclass[twocolumn]{aastex702}
\usepackage{graphicx}	
\usepackage{amsmath}	
\usepackage{amssymb}
\usepackage{color}
\usepackage{array}
\usepackage{bookmark}
\usepackage{subfigure}
\usepackage{booktabs}
\usepackage{threeparttable}
\def\beq{\begin{eqnarray}}
\def\eeq{\end{eqnarray}}

\begin{document}

\title{A statistical relation of energy injection plateaus in multi-band afterglows of gamma-ray bursts}

\author[0000-0002-6588-2652]{Xiao-Yan Li}
\affiliation{School of Physics and Mechanical and Electrical Engineering, Longyan University, Longyan, Fujian 364012, China}
\affiliation{Department of Astronomy, Xiamen University, Xiamen, Fujian 361005, China; tongliu@xmu.edu.cn}
\email{lixiaoyan@stu.xmu.edu.cn}

\author[0000-0001-8678-6291]{Tong Liu}
\affiliation{Department of Astronomy, Xiamen University, Xiamen, Fujian 361005, China; tongliu@xmu.edu.cn}
\email{tongliu@xmu.edu.cn}

\author[0000-0002-4448-0849]{Bao-Quan Huang}
\affiliation{College of Intelligent Manufacturing, Nanning University, Nanning, Guangxi 530299, China; huangbaoquan@unn.edu.cn}
\email{huangbaoquan@unn.edu.cn}

\begin{abstract}
The origin of the plateau phase in gamma-ray burst (GRB) afterglows remains under debate, with the energy injection model being one of the most competitive explanations. If the plateau is truly driven by energy injection, the average X-ray and optical luminosities during the plateau phase, $L_{\rm X, plat, ave}$ and $L_{\rm opt, plat, ave}$, should naturally be correlated. Moreover, under this scenario, the scaling relations between the luminosities during the plateau and the normal decay phase are expected to be consistent since they share the same origin, i.e., synchrotron radiation from the external forward shock. Therefore, simultaneous multi-band observations are essential to verify this mechanism. In this work, we select a sample of 47 GRBs with simultaneous plateaus in both bands. We calculate their time-averaged isotropic luminosities for the plateau and the subsequent normal decay phases. We find a moderate positive correlation $\log L_{\rm X, plat, ave}=m\log L_{\rm opt, plat, ave}+c$ with a slope $m = 0.86 \pm 0.11$ for the plateau phase. This correlation supports the energy injection origin and offers a promising diagnostic approach to test the model. Furthermore, we obtain a similar slope $m = 1.05 \pm 0.05$ for the normal decay phase, which reinforces the idea that both phases share the same physical origin. Notably, the post-plateau data exhibit a systematic downward shift in luminosity, which may indicate the cessation of the central engine.
\end{abstract}

\keywords{gamma-ray bursts (629)} 

\section{Introduction}
Gamma-ray bursts (GRBs) are observationally characterized by an intense, short-lived burst of $\gamma$-ray emission, i.e., the prompt emission, followed by a long-lasting multiband afterglow spanning from radio to X-ray frequencies. The standard fireball model has been highly successful in explaining many observed properties of GRB afterglows. In this model, the synchrotron radiation is produced by a relativistic jet propagating through the circumburst medium \citep[e.g.,][]{1997ApJ...476..232M, 1998ApJ...497L..17S}. Within this standard framework, the temporal decay index $\alpha$ and the spectral index $\beta$ (where $F_\nu \propto t^{-\alpha}\nu^{-\beta}$) are theoretically linked by specific algebraic expressions, known as closure relations, which depend on the jet dynamics, the density profile of the surrounding medium, and the cooling regime of the shock-accelerated electrons \citep[e.g.,][]{Zhang2004, 1998ApJ...497L..17S}. These relations have been widely used in the literature to constrain the physical parameters of GRB afterglows \citep[e.g.,][]{Racusin2009}.

While this model provides a successful framework for interpreting the multiband afterglows, observations by the \textit{Swift} satellite revealed diverse and complex temporal features in GRB afterglow light curves that deviate from the simple power-law decay predicted by the standard model \citep[e.g.,][]{2004ApJ...611.1005G,2005SSRv..120..165B}. In the X-ray band, a canonical light curve has been proposed, consisting of several distinct segments: an initial steep decay, a shallow decay phase (or plateau) with a typical temporal slope of $\sim -0.5$ \citep[e.g.,][]{2006ApJ...642..354Z}, a normal decay, characterized by a typical temporal slope of $\sim 1$--$1.5$, consistent with the standard afterglow model, a late steep decay following the jet-break, and superimposed X-ray flares \citep[e.g.,][]{2006ApJ...642..389N,2006ApJ...647.1213O,2006ApJ...642..354Z,2007ApJ...666.1002Z}. In the optical band, \citet{2012ApJ...758...27L} identified eight possible emission components in afterglow light curves: prompt optical flares tracking the $\gamma$-ray emission, early optical flashes associated with external reverse shocks, a plateau phase, a normal decay, a post-jet-break phase, optical flares, re-brightening features, and supernova bumps.

Among the diverse features observed in GRB afterglow light curves, the plateau phase stands out as one of the most intriguing and frequently observed phenomena, appearing across multiple bands. Its physical origin remains under debate. A variety of models have been proposed to account for this feature, including off-axis jet observations \citep[e.g.,][]{2006ApJ...641L...5E,2007ApJ...655..973K}, evolving microphysical parameters \citep[e.g.,][]{2006AA...458....7I,2006MNRAS.369..197F}, structured jet effects \citep[e.g.,][]{2006MNRAS.370.1946G,2007ApJ...656L..57J,2021ApJ...922...22L}, a two-component ejecta model \citep[e.g.,][]{2006ApJ...640L.139T,2009ApJ...690L.118Y}, a delayed jet deceleration scenario \citep{2015ApJ...806..205D}, and a precessing jet \citep[e.g.,][]{2021ApJ...916...71H,2024Univ...10..438H}. Of course, some of the models imply the modes of continuous or intermittent energy injection. Recent statistical and multi-wavelength studies have further strengthened these interpretations and provided tighter constraints on the underlying physical mechanisms \citep[e.g.,][]{2020ApJ...904...97D,2022NatCo..13.5611D,2022ApJ...924...69Y}.

One of the most promising interpretations is the energy-injection model, in which a long-lived central engine, such as a millisecond magnetar \citep[e.g.,][]{1998AA...333L..87D,2001ApJ...552L..35Z, 2013ApJ...779L..25F,2026ApJ..1000...97D}, phase transition of a supramassive fast-rotating quark star \citep[e.g.,][]{2016PhRvD..94h3010L,2018ApJ...854..104H}, or a fallback/accreting black hole (BH) central engine \citep[e.g.,][]{Kumar2008,Cannizzo2009,Cannizzo2011,2017NewAR..79....1L,Lenart2025}, continuously injects energy into the forward shock, thereby sustaining the plateau phase \citep[e.g.,][]{1998AA...333L..87D,2006ApJ...642..354Z,2006ApJ...642..389N,2006MNRAS.369.2059P,2006MNRAS.366.1357P}. In this scenario, a plateau is naturally produced provided that the energy injection satisfies certain conditions \citep{2001ApJ...552L..35Z}. Under energy injection, standard closure relations are modified to incorporate the energy injection rate $q$ \citep[e.g.,][]{2001ApJ...552L..35Z, 2006ApJ...642..354Z}, but this introduces extra parameters and degeneracies, making it challenging to uniquely constrain the physical models through individual closure relation tests. Once the injection ceases or becomes negligible, the light curve transitions to the steeper ``normal'' decay phase.

If the X-ray and optical plateaus share a common energy-injection origin, the end of the plateau phase, i.e., the temporal break time $T_{\rm b}$, is expected to be achromatic across these bands \citep[e.g.,][]{2007ApJ...670..565L}. In addition, the average luminosities during the plateau phase are naturally expected to be correlated. Furthermore, the scaling relations of the average luminosities between the plateau and the normal decay phases are expected to be consistent, as both stages originate from synchrotron radiation of the external forward shock. Specifically, the plateaus very likely originate from a refreshed forward shock\citep[e.g.,][]{2006ApJ...642..354Z, 2019ApJ...877..147K}. Therefore, comprehensive multi-wavelength observational studies are crucial to study these cross-band correlations and test the energy-injection model.

In this paper, by constructing a high-quality sample of GRBs with simultaneous, well-defined plateaus in both X-ray and optical bands, we perform a systematic analysis of their luminosity correlations to test the energy-injection model. We calculate the time-averaged X-ray and optical luminosities during the plateau phase, $L_{\rm X, plat, ave}$ and $L_{\rm opt, plat, ave}$, to examine whether a correlation exists. Furthermore, we calculate the average luminosities of the two bands of the subsequent normal decay phase, $L_{\rm X, decay, ave}$ and $L_{\rm opt, decay, ave}$ to assess whether the underlying radiation mechanism remains consistent across the light-curve break.

This paper is organized as follows. Section~\ref{sec:method} describes the sample selection and the theoretical framework of the energy-injection model. Section~\ref{sec:results} presents the main results. Section~\ref{sec:summary}  summarizes our findings and discusses their physical implications.

\section{Method} 
\label{sec:method} 
\subsection{Sample Selection} 
Following the criteria in \citet{2026ApJ...998..298L}, a plateau is defined as a light curve segment where the temporal power-law decay index $\alpha$ satisfies $-1 \leq \alpha \leq 1$ under the convention $F_\nu \propto t^{-\alpha}\nu^{-\beta}$. The light curves are modeled using either a single power-law (SPL) or a smoothly broken power-law (BPL). The SPL model is described as
\begin{equation}
F(t) = F_0 \, t^{-\alpha},
\end{equation}
where $F(t)$ represents the flux at time $t$, $F_0$ is the flux amplitude normalization, and $\alpha$ denotes the temporal decay index.

For bursts exhibiting a break in the light curve, the BPL model is applied \citep[e.g.,][]{2015ApJS..219....9W,2015ApJ...805...13L,2022NatCo..13.5611D}:
\begin{equation}
F(t) = F_0 \left[ \left( \frac{t}{T_{\rm b}} \right)^{\alpha_1 \omega} + \left( \frac{t}{T_{\rm b}} \right)^{\alpha_2 \omega} \right]^{-1/\omega}.
\label{eq:bkpl}
\end{equation}
In this expression, $T_{\rm b}$ represents the break time separating two distinct decay phases. The parameters $\alpha_1$ and $\alpha_2$ correspond to the temporal decay indices before and after the break time. The parameter $\omega$ quantifies the sharpness of the transition at the break and is fixed as $\omega = 1$ in accordance with the methodology of \citet{2026ApJ...998..298L}. The flux at the break time is related to the normalization parameter by $F_{\rm b} = 2^{-1/\omega} F_0 $. 

To investigate the correlation between X-ray and optical emissions in different evolutionary stages of GRB afterglows, we utilized the GRB sample of \citet{2026ApJ...998..298L}. Specifically, we focused on Dataset 1, which consists of 75 GRBs characterized by the simultaneous detection of plateau phases in both X-ray and optical bands. To ensure the reliability of our analysis, we further narrowed this sample down to 47 GRBs with well-constrained break times $T_{\rm b}$ in both bands. These sources have clear spectroscopic redshifts ($z$) and contain sufficient data points ($>5$) in the light curves of both bands. We directly adopt all temporal parameters (e.g., break time $T_{\rm b}$, decay indices $\alpha$, $\alpha_1$, $\alpha_2$) and spectral indices ($\beta$) from that paper.  

\begin{table*}[htbp]
\centering
\tiny
\caption{Properties of the multi-band afterglow plateau and decay phases}
\label{tab:luminosity_plateau}
\setlength{\tabcolsep}{1.2pt}
\begin{tabular}{lccccccccccccccc}
\toprule
 GRB & $z$ & $T_{\rm 90}$(s) & $T_{\rm b,X}$(ks) & $\alpha_{\rm X,1}$ & $\alpha_{\rm X,2}$ & \textbf{$\beta_{\rm X}$} & $L_{\rm X, plat, ave}$ & $L_{\rm X, decay, ave}$ & $T_{\rm b,o}$(ks) & $\alpha_{\rm o,1}$ & $\alpha_{\rm o,2}$ & \textbf{$\beta_{\rm o}$} & $L_{\rm opt, plat, ave}$ & $L_{\rm opt, decay, ave}$ & Ref\\
  &  & (s) & (ks) &  &  & \textbf{ } & ($10^{45}$\,erg/s) & ($10^{45}$\,erg/s) & (ks) &  &  & \textbf{ } & ($10^{44}$\,erg/s) & ($10^{44}$\,erg/s) & \\
\midrule
 050319  & 3.24  & 15  & 55.00  & $0.58^{+0.07}_{-0.07}$  & $1.72^{+0.11}_{-0.11}$  & \textbf{$1.01^{+0.07}_{-0.07}$}  & 121.50  & 17.72  & -  & $0.39^{+0.06}_{-0.06}$  & $1.02^{+0.04}_{-0.04}$  & \textbf{$0.74^{+0.42}_{-0.42}$}  & 7.36  & 16.58  & 1 \\
 050401  & 2.90  & 38  & 4.30  & $0.76^{+0.05}_{-0.05}$  & $1.67^{+0.11}_{-0.11}$  & \textbf{$0.79^{+0.13}_{-0.13}$}  & 2536.43  & 343.39  & -  & $0.50^{+0.08}_{-0.08}$  & $0.89^{+0.08}_{-0.08}$  & \textbf{$0.50^{+0.20}_{-0.20}$}  & 3.18  & 2.02  & 1 \\
 050408  & 1.24  & 34  & 40.70  & $0.73^{+0.15}_{-0.15}$  & $1.17^{+0.22}_{-0.22}$  & \textbf{$1.14^{+0.14}_{-0.14}$}  & 23.57  & 1.28  & -  & $0.49^{+0.01}_{-0.01}$  & $1.29^{+0.11}_{-0.11}$  & \textbf{$0.28^{+0.33}_{-0.33}$}  & 0.36  & 0.24  & 1 \\
 050416A  & 0.65  & 2.4  & 11.00  & $0.66^{+0.12}_{-0.12}$  & $0.99^{+0.02}_{-0.02}$  & \textbf{$1.07^{+0.11}_{-0.11}$}  & 9.12  & 0.14  & -  & $0.26^{+0.07}_{-0.07}$  & $1.12^{+0.12}_{-0.12}$  & \textbf{$1.30$}  & 0.24  & 0.43  & 1 \\
 050730  & 3.97  & 155  & 90.10  & $0.45^{+0.13}_{-0.13}$  & $2.64^{+0.20}_{-0.20}$  & \textbf{$1.62^{+0.04}_{-0.04}$}  & 84.14  & 81.32  & -  & $0.48^{+0.05}_{-0.05}$  & $1.47^{+0.06}_{-0.06}$  & \textbf{$0.52^{+0.05}_{-0.05}$}  & 5.15  & 17.70  & 1 \\
 050801  & 1.38  & 20  & 0.20  & $0.24^{+0.11}_{-0.11}$  & $1.18^{+0.03}_{-0.03}$  & \textbf{$0.92^{+0.17}_{-0.17}$}  & 200.98  & 2.92  & -  & $0.07^{+0.01}_{-0.01}$  & $1.20^{+0.01}_{-0.01}$  & \textbf{$0.69^{+0.34}_{-0.34}$}  & 210.54  & 19.70  & 1 \\
 050922C  & 2.20  & 5  & 3.20  & $0.97^{+0.04}_{-0.04}$  & $1.49^{+0.05}_{-0.05}$  & \textbf{$0.17^{+0.11}_{-0.11}$}  & 473.56  & 62.55  & 6.40  & $0.46^{+0.03}_{-0.03}$  & $1.50^{+0.02}_{-0.02}$  & \textbf{$0.51^{+0.05}_{-0.05}$}  & 36.60  & 10.65  & 2 \\
 051221A  & 0.65  & 2.4  & 25.10  & $0.35^{+0.08}_{-0.08}$  & $1.34^{+0.04}_{-0.04}$  & \textbf{$1.06^{+0.14}_{-0.14}$}  & 2.37  & 0.17  & -  & $0.34^{+0.07}_{-0.07}$  & $1.24^{+0.04}_{-0.04}$  & \textbf{$0.64^{+0.05}_{-0.05}$}  & 0.45  & 0.55  & 1 \\
 060206  & 4.06  & 11  & 12.50  & $0.40^{+0.09}_{-0.09}$  & $1.50^{+0.06}_{-0.06}$  & \textbf{$1.20^{+0.31}_{-0.31}$}  & 1348.68  & 31.22  & -  & $0.42^{+0.09}_{-0.09}$  & $1.43^{+0.10}_{-0.10}$  & \textbf{$0.73^{+0.05}_{-0.05}$}  & 175.66  & 69.53  & 1 \\
 060210  & 3.91  & 255  & 5.00  & $0.53^{+0.06}_{-0.06}$  & $1.30^{+0.12}_{-0.12}$  & \textbf{$1.08^{+0.08}_{-0.08}$}  & 2641.40  & 114.08  & -  & $0.53^{+0.05}_{-0.05}$  & $1.77^{+0.14}_{-0.14}$  & \textbf{$0.37^{+0.08}_{-0.08}$}  & 4.71  & 31.49  & 1 \\
 060526  & 3.22  & 13.8  & 50.10  & $0.67^{+0.08}_{-0.08}$  & $2.06^{+0.15}_{-0.15}$  & \textbf{$0.90^{+0.11}_{-0.11}$}  & 265.34  & 312.98  & -  & $0.56^{+0.10}_{-0.10}$  & $1.93^{+0.10}_{-0.10}$  & \textbf{$0.51^{+0.32}_{-0.32}$}  & 12.36  & 17.35  & 1 \\
 060607A  & 3.07  & 100  & 9.50  & $0.36^{+0.03}_{-0.03}$  & $3.10^{+0.12}_{-0.12}$  & \textbf{$0.62^{+0.06}_{-0.06}$}  & 1289.23  & 114.30  & -  & -0.93  & $4.6$  & \textbf{$0.72^{+0.27}_{-0.27}$}  & 111.11  & 1821.70  & 1 \\
 060614  & 0.13  & 102  & 44.00  & $0.11^{+0.10}_{-0.10}$  & $1.97^{+0.15}_{-0.15}$  & \textbf{$0.90^{+0.09}_{-0.09}$}  & 0.26  & 0.01  & -  & $-0.35^{+0.04}_{-0.04}$  & $1.90^{+0.12}_{-0.12}$  & \textbf{$0.47^{+0.04}_{-0.04}$}  & 0.06  & 0.12  & 1 \\
 $060708^{\rm s}$  & 9.8  & 1.92  & 8.90  & $0.60^{+0.05}_{-0.05}$  & $1.32^{+0.05}_{-0.05}$  & \textbf{$0.16^{+0.20}_{-0.20}$}  & 235.05  & 20.46  & 0.70  & $0.06^{+0.10}_{-0.10}$  & $0.85^{+0.03}_{-0.03}$  & \textbf{$0.88^{+0.05}_{-0.05}$}  & 608.82  & 980.57  & 2 \\
 060714  & 2.71  & 115  & 5.90  & $0.48^{+0.09}_{-0.09}$  & $1.34^{+0.11}_{-0.11}$  & \textbf{$1.10^{+0.19}_{-0.19}$}  & 343.67  & 12.03  & -  & $0.15^{+0.06}_{-0.06}$  & $1.04^{+0.15}_{-0.15}$  & \textbf{$0.44^{+0.04}_{-0.04}$}  & 4.00  & 46.39  & 1 \\
 060729  & 0.54  & 116  & 53.00  & $0.05^{+0.01}_{-0.01}$  & $1.45^{+0.11}_{-0.11}$  & \textbf{$1.02^{+0.04}_{-0.04}$}  & 17.79  & 0.27  & -  & $0.10^{+0.05}_{-0.05}$  & $1.40^{+0.15}_{-0.15}$  & \textbf{$0.78^{+0.03}_{-0.03}$}  & 3.95  & 0.55  & 1 \\
 060908  & 1.88  & 19.3  & 1.10  & $0.54^{+0.05}_{-0.05}$  & $1.66^{+0.05}_{-0.05}$  & \textbf{$1.13^{+0.19}_{-0.19}$}  & 853.21  & 4.96  & -  & $0.71^{+0.04}_{-0.04}$  & $1.11^{+0.09}_{-0.09}$  & \textbf{$0.24^{+0.20}_{-0.20}$}  & 18.83  & 5.06  & 1 \\
 061021  & 0.35  & 46  & 16.60  & $0.60^{+0.03}_{-0.03}$  & $1.16^{+0.02}_{-0.02}$  & \textbf{$0.98^{+0.10}_{-0.10}$}  & 7.46  & 0.08  & 88.80  & $0.65^{+0.01}_{-0.01}$  & $2.01^{+0.24}_{-0.24}$  & \textbf{$0.55^{+0.02}_{-0.02}$}  & 0.13  & 0.05  & 2 \\
 061121  & 1.31  & 81  & 7.70  & $0.34^{+0.03}_{-0.03}$  & $1.55^{+0.03}_{-0.03}$  & \textbf{$0.93^{+0.02}_{-0.02}$}  & 446.76  & 5.48  & 40.90  & $0.53^{+0.08}_{-0.08}$  & $1.26^{+0.14}_{-0.14}$  & \textbf{$0.60$}  & 1.76  & 1.62  & 3 \\
 070110  & 2.35  & 85  & 20.30  & $0.30^{+0.10}_{-0.10}$  & $5.10^{+0.30}_{-0.30}$  & \textbf{$1.08^{+0.11}_{-0.11}$}  & 234.39  & 4.97  & -  & $0.16^{+0.06}_{-0.06}$  & $1.67^{+0.12}_{-0.12}$  & \textbf{$0.55^{+0.04}_{-0.04}$}  & 2.76  & 4.97  & 1 \\
 071112C  & 0.82  & 15  & 1.50  & $0.50^{+0.08}_{-0.08}$  & $1.49^{+0.11}_{-0.11}$  & \textbf{$0.67^{+0.13}_{-0.13}$}  & 105.40  & 2.35  & -  & $0.30^{+0.07}_{-0.07}$  & $0.95^{+0.11}_{-0.11}$  & \textbf{$0.63^{+0.29}_{-0.29}$}  & 2.82  & 0.33  & 1 \\
 080310  & 2.43  & 365  & 5.10  & $0.03^{+0.06}_{-0.06}$  & $1.24^{+0.08}_{-0.08}$  & \textbf{$0.95^{+0.18}_{-0.18}$}  & 199.29  & 13.91  & -  & $0.11^{+0.01}_{-0.01}$  & $1.24^{+0.02}_{-0.02}$  & \textbf{$0.42^{+0.12}_{-0.12}$}  & 21.93  & 33.23  & 1 \\
 080319B  & 0.94  & 50  & 3.00  & $0.73^{+0.05}_{-0.05}$  & $2.73^{+0.16}_{-0.16}$  & \textbf{$0.81^{+0.07}_{-0.07}$}  & 4589.43  & 4.35  & -  & $0.93^{+0.11}_{-0.11}$  & $1.60^{+0.12}_{-0.12}$  & \textbf{$0.51^{+0.26}_{-0.26}$}  & 137.31  & 5.20  & 1 \\
 080413B  & 1.10  & 8  & 148.50  & $0.92^{+0.16}_{-0.16}$  & $1.91^{+0.23}_{-0.23}$  & \textbf{$0.94^{+0.07}_{-0.07}$}  & 10.21  & 1.19  & -  & $0.31^{+0.15}_{-0.15}$  & $1.89^{+0.22}_{-0.22}$  & \textbf{$0.25^{+0.07}_{-0.07}$}  & 1.00  & 0.31  & 1 \\
 080710  & 0.85  & 120  & 6.80  & $0.34^{+0.04}_{-0.04}$  & $1.57^{+0.14}_{-0.14}$  & \textbf{$1.00^{+0.11}_{-0.11}$}  & 26.23  & 1.00  & -  & $0.39^{+0.05}_{-0.05}$  & $1.32^{+0.11}_{-0.11}$  & \textbf{$0.80^{+0.09}_{-0.09}$}  & 22.26  & 28.74  & 1 \\
 081008  & 1.97  & 185.5  & 9.50  & $0.87^{+0.15}_{-0.15}$  & $1.68^{+0.08}_{-0.08}$  & \textbf{$0.98^{+0.11}_{-0.11}$}  & 227.38  & 4.04  & -  & $0.64^{+0.06}_{-0.06}$  & $1.60^{+0.09}_{-0.09}$  & \textbf{$0.40^{+0.23}_{-0.23}$}  & 33.60  & 10.63  & 1 \\
 $090426^{\rm s}$  & 2.6  & 1.2  & 0.20  & $0.13^{+0.02}_{-0.02}$  & $1.04^{+0.05}_{-0.05}$  & \textbf{$1.03^{+0.15}_{-0.15}$}  & 529.87  & 19.80  & -  & $0.14^{+0.09}_{-0.09}$  & $1.25^{+0.04}_{-0.04}$  & \textbf{$0.76^{+0.14}_{-0.14}$}  & 118.84  & 131.40  & 1 \\
 090529A  & 2.6  & 100  & 97.50  & $0.52^{+0.06}_{-0.06}$  & $1.09^{+0.20}_{-0.20}$  & \textbf{$0.48^{+0.67}_{-0.23}$}  & 4.46  & 2.01  & 44.53  & $0.45^{+0.03}_{-0.03}$  & $0.90^{+0.07}_{-0.07}$  & \textbf{$0.10^{+0.46}_{-0.46}$}  & 0.86  & 3.01  & 4 \\
 090618  & 0.54  & 113.2  & 45.10  & $0.93^{+0.09}_{-0.09}$  & $1.74^{+0.10}_{-0.10}$  & \textbf{$0.92^{+0.05}_{-0.05}$}  & 45.78  & 0.43  & -  & $0.76^{+0.11}_{-0.11}$  & $1.53^{+0.11}_{-0.11}$  & \textbf{$0.50^{+0.05}_{-0.05}$}  & 2.54  & 0.10  & 1 \\
 091018  & 0.97  & 4.4  & 0.30  & $-0.17^{+0.26}_{-0.26}$  & $1.25^{+0.03}_{-0.03}$  & \textbf{$0.92^{+0.03}_{-0.03}$}  & 794.15  & 4.42  & 125.10  & $0.91^{+0.01}_{-0.01}$  & $2.84^{+0.32}_{-0.32}$  & \textbf{$0.61^{+0.02}_{-0.02}$}  & 0.79  & 0.63  & 3 \\
 091029  & 2.75  & 39.2  & 20.80  & $0.32^{+0.09}_{-0.09}$  & $1.35^{+0.08}_{-0.08}$  & \textbf{$1.12^{+0.08}_{-0.08}$}  & 126.55  & 11.23  & -  & $0.48^{+0.06}_{-0.06}$  & $1.34^{+0.09}_{-0.09}$  & \textbf{$0.49^{+0.12}_{-0.12}$}  & 10.75  & 11.35  & 1 \\
 091127  & 0.49  & 7.1  & 35.30  & $0.96^{+0.05}_{-0.05}$  & $1.59^{+0.12}_{-0.12}$  & \textbf{$0.68^{+0.11}_{-0.11}$}  & 102.96  & 0.11  & -  & $0.55^{+0.11}_{-0.11}$  & $1.50^{+0.11}_{-0.11}$  & \textbf{$0.18$}  & 2.36  & 0.16  & 1 \\
 100219A  & 4.67  & 18.8  & 1.80  & $0.54^{+0.07}_{-0.07}$  & $1.65^{+0.15}_{-0.15}$  & \textbf{$0.69^{+0.23}_{-0.23}$}  & 898.38  & 300.91  & -  & $0.74^{+0.08}_{-0.08}$  & $1.91^{+0.12}_{-0.12}$  & \textbf{$0.56$}  & 25.94  & 43.14  & 1 \\
 100418A  & 0.62  & 7  & 90.10  & $-0.12^{+0.03}_{-0.03}$  & $1.57^{+0.11}_{-0.11}$  & \textbf{$1.04^{+0.29}_{-0.29}$}  & 0.83  & 0.10  & -  & $0.11^{+0.01}_{-0.01}$  & $1.60^{+0.10}_{-0.10}$  & \textbf{$0.98^{+0.09}_{-0.09}$}  & 0.33  & 0.58  & 1 \\
 100621A  & 0.54  & 63.6  & 80.50  & $0.68^{+0.03}_{-0.03}$  & $1.69^{+0.09}_{-0.09}$  & \textbf{$0.37^{+0.03}_{-0.03}$}  & 8.34  & 0.31  & 40.00  & $0.22^{+0.02}_{-0.02}$  & $2.41^{+0.18}_{-0.18}$  & \textbf{$0.78^{+0.09}_{-0.09}$}  & 4.43  & 1.90  & 3 \\
 100814A  & 1.44  & 177  & 215.30  & $0.50^{+0.02}_{-0.02}$  & $2.38^{+0.09}_{-0.09}$  & \textbf{$0.84^{+0.02}_{-0.02}$}  & 19.86  & 1.17  & 391.40  & $0.09^{+0.01}_{-0.01}$  & $3.90^{+0.15}_{-0.15}$  & \textbf{$0.41^{+0.04}_{-0.04}$}  & 1.23  & 0.48  & 3\\
 110213A  & 1.46  & 48  & 3.10  & $-0.19^{+0.06}_{-0.06}$  & $1.90^{+0.03}_{-0.03}$  & \textbf{$0.04^{+0.02}_{-0.02}$}  & 334.92  & 12.82  & 15.20  & $0.20^{+0.01}_{-0.01}$  & $2.01^{+0.02}_{-0.02}$  & \textbf{$0.90^{+0.07}_{-0.07}$}  & 106.98  & 237.16  & 3 \\
 110715A  & 0.82  & 13  & 0.10  & $-0.80^{+0.32}_{-0.32}$  & $1.00^{+0.01}_{-0.01}$  & \textbf{$0.85^{+0.02}_{-0.02}$}  & 1575.84  & 9.34  & 374.90  & $0.52^{+0.01}_{-0.01}$  & $2.87^{+0.30}_{-0.30}$  & \textbf{$0.63^{+0.28}_{-0.28}$}  & 0.04  & 0.12  & 3 \\
 140518A  & 4.70  & 60.5  & 4.00  & $0.45^{+0.05}_{-0.05}$  & $1.93^{+0.13}_{-0.13}$  & \textbf{$0.91^{+0.12}_{-0.12}$}  & 496.49  & 478.40  & -  & $-0.02^{+0.05}_{-0.05}$  & $1.06^{+0.07}_{-0.07}$  & \textbf{$0.22^{+0.10}_{-0.34}$}  & 13.38  & 215.71  & 4 \\
 140703A  & 3.14  & 67.1  & 20.40  & $0.91^{+0.06}_{-0.06}$  & $2.35^{+0.12}_{-0.12}$  & \textbf{$0.81^{+0.09}_{-0.09}$}  & 821.13  & 78.66  & -  & $-0.10^{+0.07}_{-0.07}$  & $1.65^{+0.10}_{-0.10}$  & \textbf{$0.81^{+0.07}_{-0.07}$}  & 15.11  & 215.54  & 4 \\
 151027A  & 0.81  & 129.7  & 9.81  & $0.37^{+0.02}_{-0.03}$  & $1.99^{+0.02}_{-0.00}$  & \textbf{$0.98^{+0.06}_{-0.06}$}  & 337.49  & 5.31  & -  & $-0.30^{+0.06}_{-0.07}$  & $1.76^{+0.02}_{-0.01}$  & \textbf{$0.80^{+0.10}_{-0.10}$}  & 46.89  & 1.63  & 5\\
 170202A  & 3.65  & 46.2  & 2.25  & $-0.05^{+0.09}_{-0.09}$  & $1.14^{+0.05}_{-0.05}$  & \textbf{$0.98^{+0.13}_{-0.12}$}  & 3263.41  & 101.69  & 4.87  & $0.92^{+0.02}_{-0.02}$  & $0.75^{+0.05}_{-0.05}$  & \textbf{$0.47^{+0.12}_{-0.65}$}  & 60.34  & 108.55  & 4 \\
 170607A  & 0.56  & 23  & 15.40  & $0.35^{+0.04}_{-0.04}$  & $0.98^{+0.02}_{-0.02}$  & \textbf{$0.94^{+0.07}_{-0.07}$}  & 14.84  & 0.94  & -  & $0.38^{+0.02}_{-0.02}$  & $0.79^{+0.12}_{-0.12}$  & \textbf{$0.90^{+0.07}_{-0.07}$}  & 2.27  & 7.11  & 4 \\
 180325A  & 2.25  & 94.1  & 7.60  & $0.82^{+0.01}_{-0.01}$  & $2.34^{+0.05}_{-0.05}$  & \textbf{$0.79^{+0.09}_{-0.09}$}  & 1885.40  & 69.83  & -  & $0.05^{+0.01}_{-0.01}$  & $1.52^{+0.04}_{-0.04}$  & \textbf{$0.45^{+0.01}_{-0.01}$}  & 3.97  & 9.46  & 4 \\
 190106A  & 1.86  & 76.8  & 16.00  & $0.36^{+0.00}_{-0.00}$  & $1.3$  & \textbf{$0.78^{+0.03}_{-0.03}$}  & 328.07  & 17.20  & 77.00  & $0.63^{+0.00}_{-0.00}$  & $1.29$  & \textbf{$0.78^{+0.03}_{-0.03}$}  & 11.24  & 16.99  & 6 \\
 210731A  & 1.25  & 22.5  & 23.33  & $0.99^{+0.16}_{-0.16}$  & $1.84^{+0.04}_{-0.04}$  & \textbf{$1.00^{+0.11}_{-0.11}$}  & 71.24  & 4.72  & 22.46  & $0.44^{+0.62}_{-0.62}$  & $1.69^{+0.19}_{-0.19}$  & \textbf{$-0.81^{+0.05}_{-0.05}$}  & 3.48  & 1.92  & 7 \\
 210905A  & 6.32  & 24  & 61.10  & $0.73^{+0.05}_{-0.05}$  & $1.09^{+0.04}_{-0.04}$  & \textbf{$0.86^{+0.13}_{-0.13}$}  & 938.39  & 85.73  & -  & $0.66^{+0.04}_{-0.04}$  & $0.94^{+0.02}_{-0.02}$  & \textbf{$0.60^{+0.04}_{-0.04}$}  & 74.47  & 39.63  & 4 \\
\bottomrule
\end{tabular}
\begin{tablenotes}
\footnotesize
\begingroup
\sloppy
\item Notes: SGRBs are marked with a superscript ``$\rm s$''; ``-'' means that the optical break time is the same as that in the X-rays. \\[-1.0em]
\item References: (1) \citet{2015ApJS..219....9W};(2) \citet{2015ApJ...805...13L}; (3) \citet{2023AA...675A.117R};  (4) this work; (5) \citet{2017AA...598A..23N}; (6) \citet{2023ApJ...948...30Z}; (7) \citet{2023AA...671A.116D}.%
\endgroup
\end{tablenotes}
\end{table*}

\subsection{Luminosity of the Plateau and Normal Decay Phase}
While many studies utilized the break luminosity for scaling relations \citep[e.g.,][]{2014ApJ...785...74L,2016ApJ...825L..20D}, we employ the time-averaged luminosity. This provides a self-consistent comparison between the plateau and the subsequent normal decay phase, which lacks a unique break epoch. Physically, the time-averaged luminosity also better represents the overall energy budget of the sustained energy injection rather than an instantaneous state.

For each phase, the isotropic radiated energy is derived by integrating the fitted temporal power-law function of the light curve over the corresponding time interval, from which the time-averaged luminosity is then obtained. Specifically, for the X-ray afterglows, assuming quasi-isotropic emission, the isotropic X-ray energy of the plateau phase $E_{\rm X, plat, iso}$ is obtained by integrating the observed luminosity over the time interval from the start of the plateau $t_{\rm s}$ to the break time $T_{\rm b}$:
\begin{equation} 
E_{\rm X, plat, iso} = \frac{4\pi d_{L}^{2} k_{\rm X}}{1+z}  \int_{t_{\rm s}}^{T_{\rm b}}  F_{\rm {b,X}} (\frac{t}{T_{\rm {b,X}}})^{-\alpha_{\rm X,1}} dt,
\label{eq:Ex_iso}
\end{equation}
where $F_{\rm {b,X}}$ is the fitted X-ray flux at the break time $T_{\rm {b,X}}$, and $\alpha_{\rm X,1}$ is the temporal decay index of the plateau phase. The $k$-correction factor is given by $k_{\rm X} = (1+z)^{\beta_{\rm X} - 1}$, where $\beta_{\rm X}$ is the X-ray spectral index. Following \citet{2015ApJS..219....9W}, we define the start time as $t_{\rm s}=\max\,(T_{90}, 60\,\mathrm{s})$ and the end time as the break time $T_{\rm b}$ determined from the light curve fitting. The total isotropic energy is derived by directly integrating this best-fit temporal function of the light curve over the plateau phase duration.

The average X-ray luminosity, $L_{\rm X, plat, ave}$, is then calculated by dividing the total energy by the duration of the plateau in the source frame:
\begin{equation}
L_{\rm X, plat, ave} = \frac{E_{\rm X, plat, iso}}{T_{\rm {b,X}} - t_{\rm {s,X}}}.
\label{eq:Lx_ave}
\end{equation}

We apply the same methodology to the optical afterglows to derive the optical isotropic energy $E_{\rm opt, plat, iso}$ and average luminosity $L_{\rm opt, plat, ave}$. We utilize the flux data from the R-band. The optical energy is calculated as:
\begin{equation} 
E_{\rm opt, plat, iso} =\frac{4\pi d_{L}^{2} k_{\rm opt}}{1+z} \int_{t_{\rm s}}^{T_{\rm {b,o}}} F_{\rm {b,o}} (\frac{t}{T_{\rm {b,o}}})^{-\alpha_{\rm o,1}} dt,
\label{eq:Eopt_iso}
\end{equation}
where $k_{\rm opt} = (1+z)^{\beta_{\rm opt} - 1}$ is the optical $k$-correction factor. Accordingly, the average optical luminosity is $L_{\rm opt, plat, ave} = E_{\rm opt, plat, iso} / (T_{\rm {b,o}} - t_{\rm {s,o}})$.

To investigate the origin of the plateau, we also calculate the average luminosity for the normal decay phase following the plateau $t > T_{\rm b}$. The isotropic energy of the normal decay phase, $E_{\rm decay, iso}$, is obtained by numerically integrating the flux over a constant source rest-frame duration $\Delta t_{\rm min}$ starting from the break time $T_{\rm b}$:
\begin{equation}
E_{\rm decay, iso} = \int_{T_{\rm b}}^{T_{\rm b} + \Delta t_{\rm min}} \frac{4\pi d_{L}^{2} \, k(z) \, (\frac{t}{T_{\rm b}})^{-\alpha_{\rm 2}}F_{\rm b}(t)}{1+z} dt,
\label{eq:E_decay}
\end{equation}
where $F_{\rm b}(t)$ represents the observed flux in the corresponding band, $k$ is the $k$-correction factor, and we set $\Delta t_{\rm min} = \Delta t_{\rm X} = 5.52\,\mathrm{ks}$ for X-rays and $\Delta t_{\rm min} = \Delta t_{\rm opt} = 15.17\,\mathrm{ks}$ for optical, which correspond to the shortest observed decay durations in our sample to avoid selection biases. To ensure consistency with the definition of the normal decay phase ($\alpha \sim 1$--$1.5$), we have verified that for all sources in our sample, the integration time interval $\Delta t_{\rm min}$ starting from $T_{\rm b}$ remains strictly within the light-curve segment dominated by the normal decay phase, avoiding any late-time jet breaks or additional complex emission components.

Consequently, the average luminosity of the decay phase is derived by dividing the total isotropic energy by the rest-frame duration:
\begin{equation}
L_{\rm decay, ave} = \frac{E_{\rm decay, iso}}{\Delta t_{\rm min}}.
\label{eq:L_decay_ave}
\end{equation}

\subsection{Energy Injection in the Forward Shock}
The plateau phase observed in GRB afterglows is frequently interpreted within the framework of the energy-injection model. This scenario postulates the existence of a long-lived central engine, such as a spinning-down magnetar or a fallback accretion BH, that continuously replenishes energy into the decelerating forward shock. The luminosity evolution of this central engine is typically parameterized as a power-law \citep[e.g.,][]{2001ApJ...552L..35Z,2015ApJS..219....9W}:
\begin{equation}
L(t) = L_{0}\left(\frac{t}{t_{0}}\right)^{-q},
\label{eq:L_t}
\end{equation}
where $t_{0}$ represents the characteristic timescale at which the external shock begins to decelerate, and $L_{0}$ is the corresponding luminosity at $t_{0}$. The parameter $q$ denotes the luminosity injection index. Effective energy injection, capable of altering the blast wave dynamics and producing a plateau, requires $q < 1$. The case of $q=0$ corresponds to a constant power injection.

We adopt this flexible power-law prescription to phenomenologically represent a broad class of central engines (including both magnetars and hyperaccreting BHs) without assuming a specific physical model. Although detailed time-dependent models predict different evolution histories, the global luminosity scalings in Equations~(\ref{LXtheory}) and (\ref{Lotheory}) remain robust, as they are governed by synchrotron shock physics and primarily determined by the total kinetic energy of the blast wave rather than the detailed injection history.

Within the framework of the standard external forward shock model, we assume the blast wave expands into a constant-density interstellar medium (ISM) and radiates via synchrotron emission. In the slow cooling regime (where the cooling frequency $\nu_{\rm c}$ is higher than the minimum injection frequency $\nu_{\rm m}$), and assuming the observing frequencies lie above the self-absorption frequency (i.e., $\nu_{\rm a} < \min(\nu_{\rm m}, \nu_{\rm c})$ and $\min(\nu_{\rm X}, \nu_{\rm opt}) > \nu_{\rm m}$), the temporal evolution of the characteristic frequencies and the peak flux can be expressed as \citep[e.g.,][]{2013NewAR..57..141G}
\begin{equation}
\nu_m = 1.37 \times 10^{18} \, {\rm Hz} \, \hat{z}^{q/2} E_{52}^{1/2} \epsilon_{e,-1}^{2} \epsilon_{B,-2}^{1/2} t^{-1-q/2},  \label{eq:num} 
\end{equation}
\begin{equation}
\nu_c = 9.2 \times 10^{18} \, {\rm Hz} \, \hat{z}^{-q/2} E_{52}^{-1/2} n_{0,0}^{-1} \epsilon_{B,-2}^{-3/2} t^{-1+q/2}, \label{eq:nuc} 
\end{equation}
and
\begin{equation}
F_{\nu,\max} = 1.1 \times 10^4 \, \mu{\rm Jy} \, \hat{z}^{q} E_{52} n_{0,0}^{1/2} \epsilon_{B,-2}^{1/2} D_{28}^{-2} t^{1-q}, \label{eq:fmax}
\end{equation}
where $E_{52}$ is the isotropic energy in units of $10^{52}$ erg, $n_{0,0}$ is the medium density in units 1 cm$^{-3}$, $D_{28}$ is the luminosity distance in $10^{28}$ cm, and $\epsilon_e$ and $\epsilon_B$ are the shock microphysics parameters. The factor $\hat{z}$ accounts for the $(1+z)$ cosmological corrections in the observer frame.

The spectral energy distribution (SED) of the afterglow in this regime is described by a broken power-law function \citep[e.g.,][]{1998ApJ...497L..17S}:
\begin{equation}
\label{Fnu_case}
F_{\nu} =
\begin{cases}
\left( \frac{\nu}{\nu_{\rm m}} \right)^{1/3} F_{\nu, \max}, & \nu < \nu_{\rm m} \\
\left( \frac{\nu}{\nu_{\rm m}} \right)^{-(p-1)/2} F_{\nu, \max}, & \nu_{\rm m} < \nu < \nu_{\rm c} \\
\left( \frac{\nu_{\rm c}}{\nu_{\rm m}} \right)^{-(p-1)/2} \left( \frac{\nu}{\nu_{\rm c}} \right)^{-p/2} F_{\nu, \max}, & \nu > \nu_{\rm c}
\end{cases}
\end{equation}
where $p$ is the electron spectral distribution index.

To directly compare the energy outputs, we convert the fluxes into luminosities using the standard relation:
\begin{equation}
L = 4\pi d_L^2(z) k\, F,
\end{equation}
where $d_L(z)$ is the luminosity distance corresponding to the redshift $z$. For the X-ray band, we calculate the time-averaged luminosity $L_{\rm X, ave}$ over the energy range of $0.3$--$10$ keV. For the optical band, we calculate the time-averaged luminosity $L_{\rm opt, ave}$ by integrating over the R-band filter passband. It should be noted that the $k$-correction factor is omitted in the following analytical derivation for simplicity. Including the $k$-correction would introduce an additional redshift-dependent normalization term in the $L_X-L_{\rm opt}$ relation, but would not modify the predicted power-law slope $m$.

Actually, the linear slope $m$ in the $\log L_{\rm X, ave} - \log L_{\rm opt, ave}$ plane can be analytically predicted from the standard synchrotron scalings. In the most typical spectral regime for early afterglows, the optical band lies in the regime $\nu_m < \nu_{\rm opt} < \nu_c$, while the X-ray band lies above the cooling frequency $\nu_{\rm X} > \max(\nu_m, \nu_c)$ \citep[e.g.,][]{2006ApJ...642..354Z}. According to Equations~(\ref{eq:num}) - (\ref{Fnu_case}), the dependencies of the multi-band luminosities on the macroscopic parameters are given by
\begin{equation}
\label{LXtheory}
L_{\rm X, ave} \propto E_{\rm iso}^{(p+2)/4} n_0^{0} \epsilon_e^{p-1} \epsilon_B^{(p-2)/4},
\end{equation}
and
\begin{equation}
\label{Lotheory}
L_{\rm opt, ave} \propto E_{\rm iso}^{(p+3)/4} n_0^{1/2} \epsilon_e^{p-1} \epsilon_B^{(p+1)/4}.
\end{equation}

It is worth noting that the power-law dependencies on the energy $E_{\rm iso}$ and the density $n_0$ are theoretically independent of the energy injection index $q$. While $q$ governs the temporal evolution of an individual burst, the global luminosity scale across a GRB ensemble is primarily dictated by their intrinsic energy spread. Therefore, if the broad dispersion of GRB luminosities is predominantly driven by the vast intrinsic diversity in their isotropic kinetic energy $E_{\rm iso}$, the theoretical slope is analytically predicted to be $m_E = (p+2)/(p+3)$ for both the plateau phase ($q=0$) and the normal decay phase ($q=1$). For a typical electron spectral index $p \approx 2.3$, this yields an expected energy-driven slope of $m_E \approx 0.81$. 

\begin{figure*}
\centering
\includegraphics[width=0.42\textwidth]{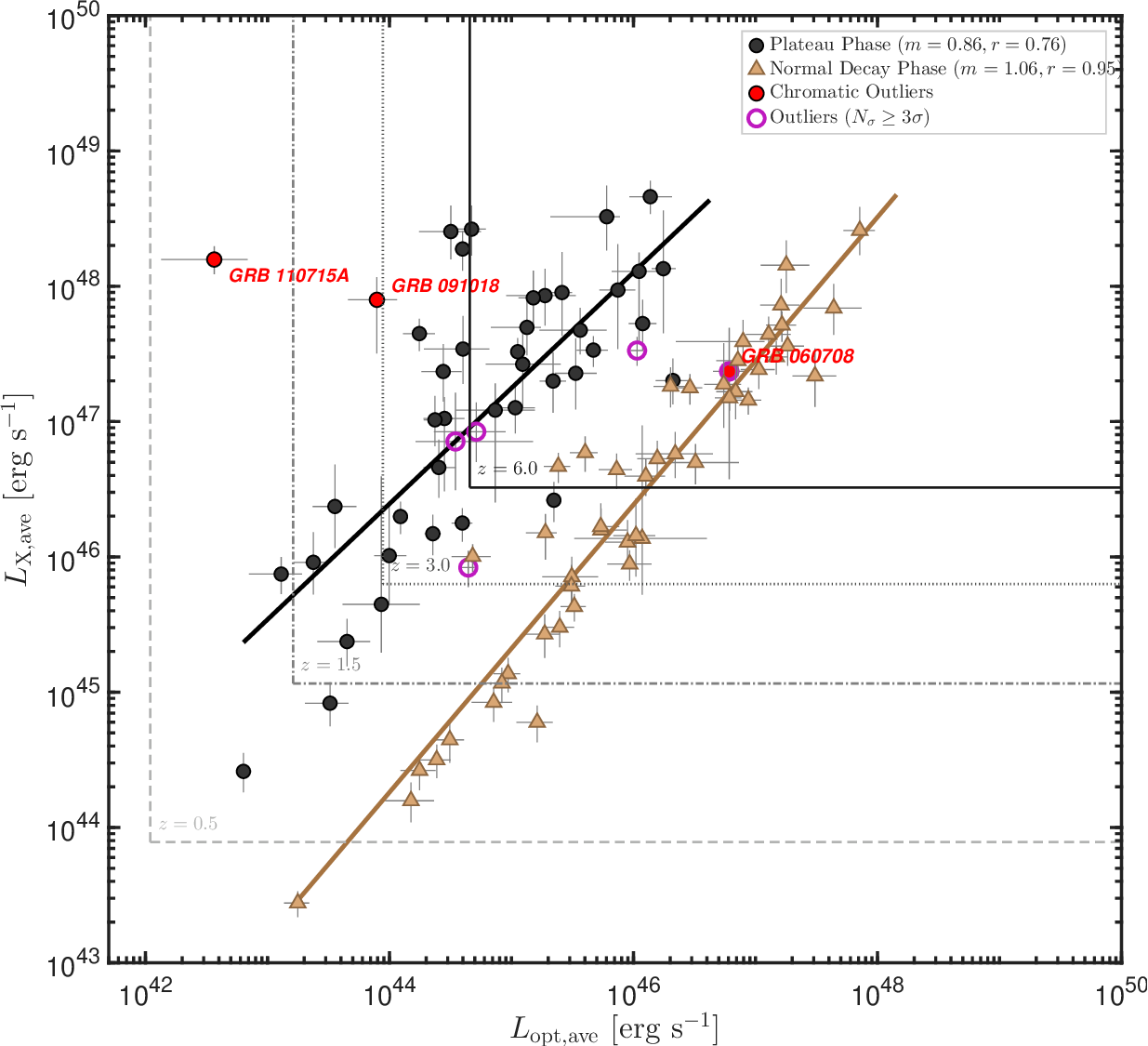}
\includegraphics[width=0.415\textwidth]{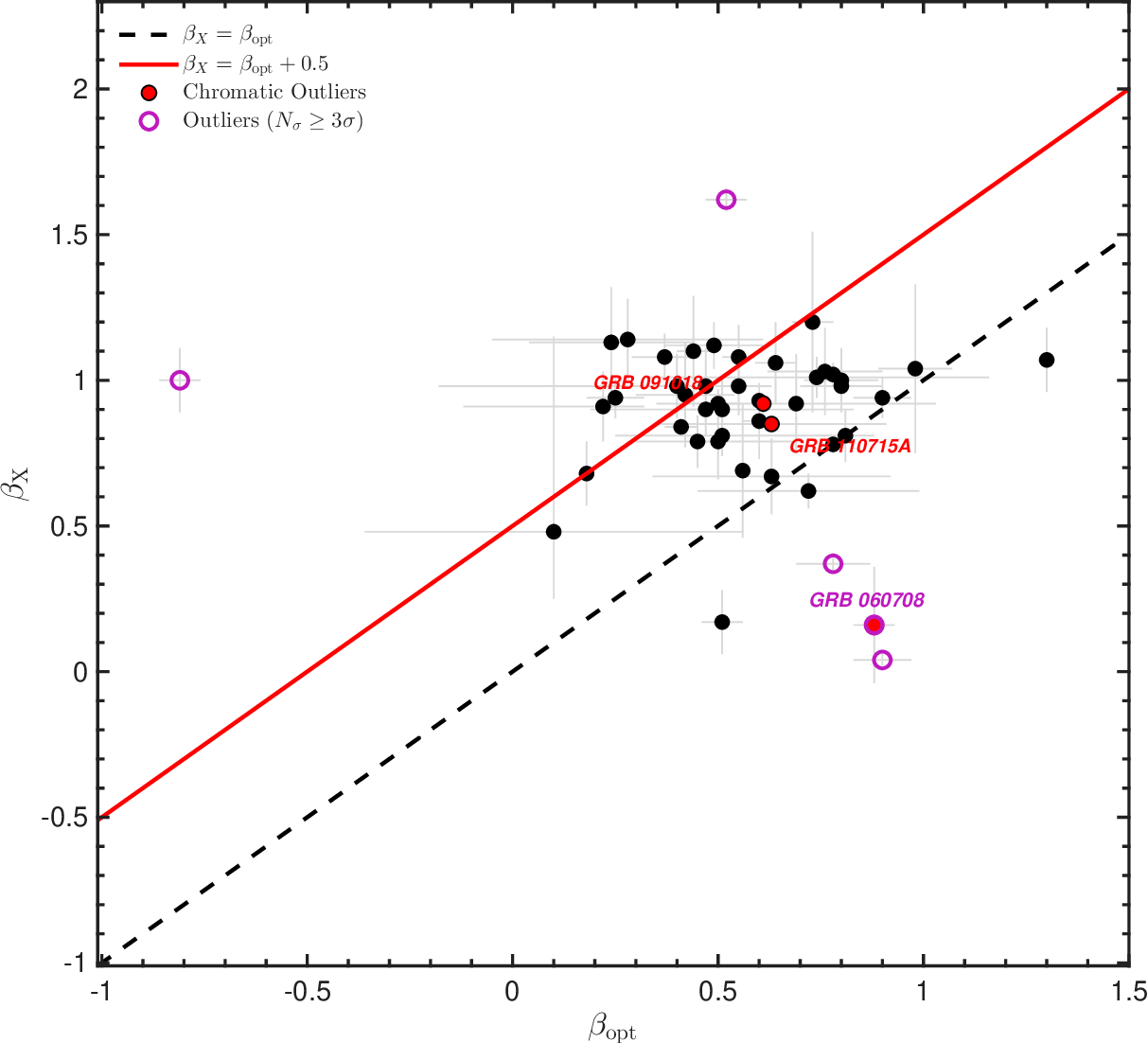}
\caption{Isotropic average luminosity and spectral index correlations for the multi-band afterglows. (a) Isotropic X-ray average luminosity $L_{\rm X, ave}$ versus optical average luminosity $L_{\rm opt, ave}$. Black circles and brown triangles denote the average luminosities during the plateau and normal decay phases, respectively. The black and brown solid lines represent the best-fit power-law relations for each phase. The chromatic plateau data points (marked in red) were excluded from the fitting process. The dashed, dash-dotted, dotted, and solid gray L-shaped lines represent the empirical selection limit boundaries for redshifts $z = 0.5$, $1.5$, $3.0$, and $6.0$, respectively (derived from the empirical plateau flux limits $F_{\rm X, lim} = 8.15 \times 10^{-14} \text{ erg s}^{-1} \text{ cm}^{-2}$ and $F_{\rm opt, lim} = 1.14 \times 10^{-15} \text{ erg s}^{-1} \text{ cm}^{-2}$). (b) X-ray spectral index $\beta_{\rm X}$ versus optical spectral index $\beta_{\rm opt}$ for the 47 GRBs in our sample. The black dashed line and red solid line represent the theoretical relations $\beta_{\rm X} = \beta_{\rm opt}$ and $\beta_{\rm X} = \beta_{\rm opt} + 0.5$, respectively.}
\label{MyFig1}
\end{figure*}

\section{results}
\label{sec:results} 
\subsection{Correlation of Optical and X-ray average Luminosities}
Figure~\ref{MyFig1}(a) displays the average luminosity distribution of our sample in the X-ray $L_{\rm X, ave}$ and optical $L_{\rm opt, ave}$ bands. The black circles represent the plateau phase, while the brown triangles denote the normal decay phase. The light gray symbols indicate the outliers that were identified and excluded from the linear regression analysis to ensure a more robust estimation of the correlation. The names of these specific GRBs are also labeled in the figure for reference.

To ensure a robust and reliable statistical correlation, several data points were excluded from the linear regression analysis. Specifically, GRB~060708, 091018, and 110715A were omitted because their plateau phases in the X-ray and optical bands were chromatic. A chromatic plateau refers to a plateau that appears in only a single band or breaks at significantly different epochs across bands \citep[e.g.,][]{2006MNRAS.369.2059P, 2007ApJ...670..565L, Oates2011}. In this work, we classify a plateau as chromatic if the X-ray and optical break times differ by more than one order of magnitude, i.e., $|\log(T_{\rm b,X} / T_{\rm b,o})| \geq 1$. Besides, GRB~210731A was also excluded from the fitting process. These points represent significant outliers that deviate substantially from the distribution of the bulk population. By removing these extreme deviations, we achieve a more accurate representation of the underlying physical correlation between $L_{\rm X, ave}$ and $L_{\rm opt, ave}$ for the majority of the GRB sample. The resulting fit (solid lines) thus reflects the intrinsic properties of the typical energy injection process.

As shown in Figure~\ref{MyFig1}(a), a moderate positive correlation between  X-ray $L_{\rm X, ave}$ and $L_{\rm opt, ave}$ is revealed in both evolutionary phases. The ordinary least squares linear regression in logarithmic space yields a moderate positive correlation $\log L_{\rm X, plat, ave}=m\log L_{\rm opt, plat, ave}+c$ with a slope $m = 0.86 \pm 0.11$ for the plateau phase, with a Pearson correlation coefficient $r=0.76$. Similarly, for the post-plateau normal decay phase, we obtain a best-fit slope of $m = 1.05 \pm 0.05$ with $r=0.95$. To assess the impact of potential outlier bias, we also apply the non-parametric Theil-Sen estimator with bootstrap resampling, which yields $m_{\rm TS} = 0.82 \pm 0.12$ for the plateau phase and $m_{\rm TS} = 1.10 \pm 0.04$ for the decay phase. These estimates are consistent with the least-squares slopes ($0.86 \pm 0.11$ and $1.05 \pm 0.05$). Furthermore, the slope difference between the two phases ($\Delta m = 0.19 \pm 0.12$) corresponds to a $1.58\sigma$ deviation ($< 2\sigma$), suggesting that the two evolutionary phases share statistically consistent scaling slopes. These similar statistical correlations suggest that the multi-band emissions during these phases share a common physical origin. 

Theoretically, the correlation during the plateau phase is a natural consequence of the multiband synchrotron emission. Since both optical and X-ray photons originate from the same forward shock produced by the jet, their luminosities are governed by the same intrinsic parameters. Variations in the parameters such as isotropic energy $E_{\rm iso}$ and ambient density $n_0$ shift the overall luminosity scale up or down. As these parameters increase, the emission in both bands rises simultaneously, naturally creating the observed positive correlation.

This theoretical invariance explains why our empirical best-fit slopes for both evolutionary phases ($m = 0.86$ and $m=1.05$) are similar. Both slopes are slightly shallower than the pure $E_{\rm iso}$-driven prediction ($m_E \approx 0.81$). This deviation perfectly aligns with the theoretical framework: as shown in Equations~(\ref{LXtheory}) and (\ref{Lotheory}), the variation in the circumburst medium density $n_0$ introduces a purely horizontal scatter in the diagram ($L_{\rm X,ave} \propto n_0^0$ while $L_{\rm opt,ave} \propto n_0^{1/2}$), which mathematically flattens the overall linear regression slope. 

To assess the potential influence of selection effects on our results, we estimate the empirical sensitivity limits of the instruments employed for our sample. The minimum time-averaged fluxes detected during the plateau phase are $F_{\rm X, lim} = 8.15 \times 10^{-14} \text{ erg s}^{-1}\text{ cm}^{-2}$ in the X-ray band and $F_{\rm opt, lim} = 1.14 \times 10^{-15} \text{ erg s}^{-1} \text{ cm}^{-2}$ in the optical band. Due to the cosmological distance effect, these flux limits correspond to redshift-dependent minimum observable luminosities: $L_{\rm min}(z) = 4\pi d_L^2(z) F_{\rm lim}$, where $d_L(z)$ is the luminosity distance calculated in a flat $\Lambda$CDM cosmology ($H_0 = 70 \text{ km s}^{-1}\text{ Mpc}^{-1}$, $\Omega_m = 0.3$, $\Omega_\Lambda = 0.7$). In Figure~\ref{MyFig1}(a), we plot these selection limits as L-shaped boundaries for several representative redshifts ($z$ = 0.5, 1.5, 3.0, and 6.0). For each redshift, the unobservable region lies below or to the left of the corresponding boundary. We find that the observed data points at various redshifts lie well within their respective observable regions (i.e., above the redshift-dependent thresholds). This suggests that the observed correlation is physical rather than an artifact of instrument detection thresholds. Note that these flux limits represent the thresholds for detecting the plateau phase; for the subsequent normal decay phase, the instruments can follow the light curves down to much lower flux levels, which explains why the decay phase data points can lie below these plateau limits. We note that these limits are empirical thresholds derived from the sample rather than the intrinsic instrumental sensitivities.

To test whether the optical and X-ray emissions indeed originate from the same synchrotron component, we compare their spectral indices in Figure~\ref{MyFig1}(b). In the standard afterglow model, if both bands lie on the same spectral segment, one expects $\beta_{\rm X}=\beta_{\rm opt}$. Alternatively, if the cooling frequency is located between the optical and X-ray bands, the model predicts $\beta_{\rm X}=\beta_{\rm opt}+0.5$ \citep{1998ApJ...497L..17S}. We find that the majority of the bursts cluster around these two relations, supporting a common synchrotron origin for the optical and X-ray emissions. A few outliers deviate significantly from the expected relations, which may indicate more complex physical conditions, such as spectral evolution, additional emission components, or uncertainties in the spectral measurements \citep[e.g.,][]{2006MNRAS.369.2059P, Oates2011}. The overall distribution is broadly consistent with the expectations of the standard synchrotron afterglow model, although the presence of several outliers suggests that a common origin may not apply to every burst in the sample.

To quantitatively evaluate spectral deviations, we define spectral closure outliers as bursts whose spectral indices depart from the theoretical boundaries ($\beta_{\rm opt} \le \beta_{\rm X} \le \beta_{\rm opt} + 0.5$) by $N_\sigma = \Delta \beta / \sqrt{\sigma_{\rm X}^2 + \sigma_{\rm opt}^2} \ge 3 \sigma$. In Figure~\ref{MyFig1}(b), these spectral outliers are highlighted as purple symbols. Correspondingly, their positions are also explicitly labeled in the luminosity correlation plot in Figure~\ref{MyFig1}(a). Notably, GRB~060708 represents a unique dual outlier, which is highlighted as a red circle with a purple border to reflect its hybrid nature that is both a chromatic plateau event and a spectral closure outlier. Interestingly, while these spectral outliers deviate from the simple closure relations in Figure~\ref{MyFig1}(b), some of them still closely follow the main empirical trend of the $L_{\rm X, ave} - L_{\rm opt, ave}$ correlation in Figure~\ref{MyFig1}(a) within the general sample dispersion. This phenomenon indicates that local spectral fluctuations may not significantly alter the overall energy scaling governed by the forward shock energy injection. Consequently, Figures~\ref{MyFig1}(a) and \ref{MyFig1}(b) complement each other well: the luminosity correlation demonstrates the intrinsic energy scaling of the bulk population, whereas the spectral closure analysis helps distinguish local radiative complexities from global engine dynamics.

We note that the assumption of a common origin for the optical and X-ray afterglows may not apply to every individual burst. For example, Ronchini et al. (2023) reported that approximately one-third of their sample showed evidence for distinct emission origins in the two bands. In the context of the present work, we focus on the statistical properties of the luminosity correlations rather than on detailed modeling of individual events. Although a subset of GRBs may involve additional emission components, the overall correlations found here suggest that a common physical mechanism, such as energy injection into the external forward shock, dominates the bulk behavior of the sample.

\begin{figure*}
\centering
\includegraphics[width=0.3\textwidth]{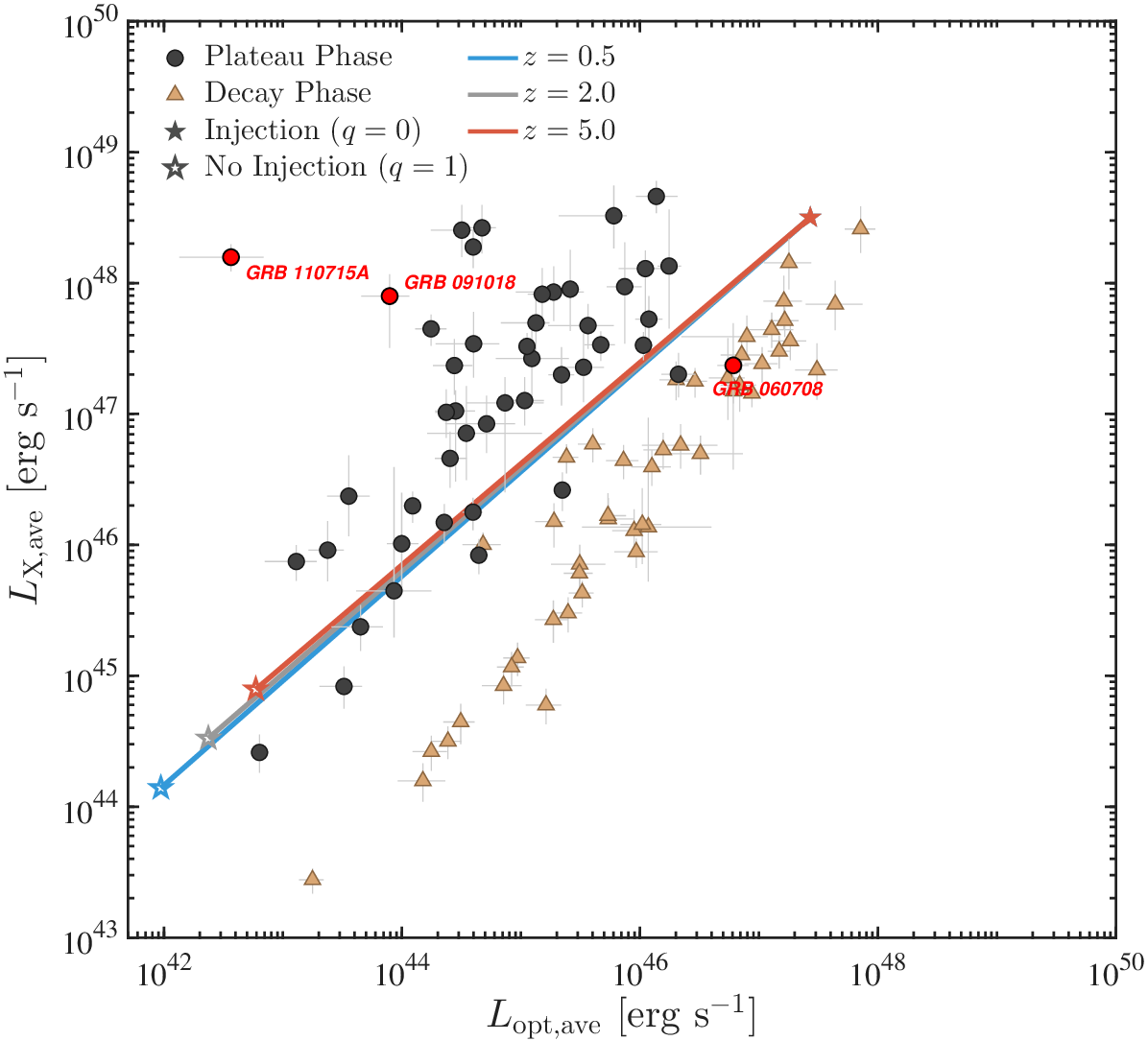}
\includegraphics[width=0.3\textwidth]{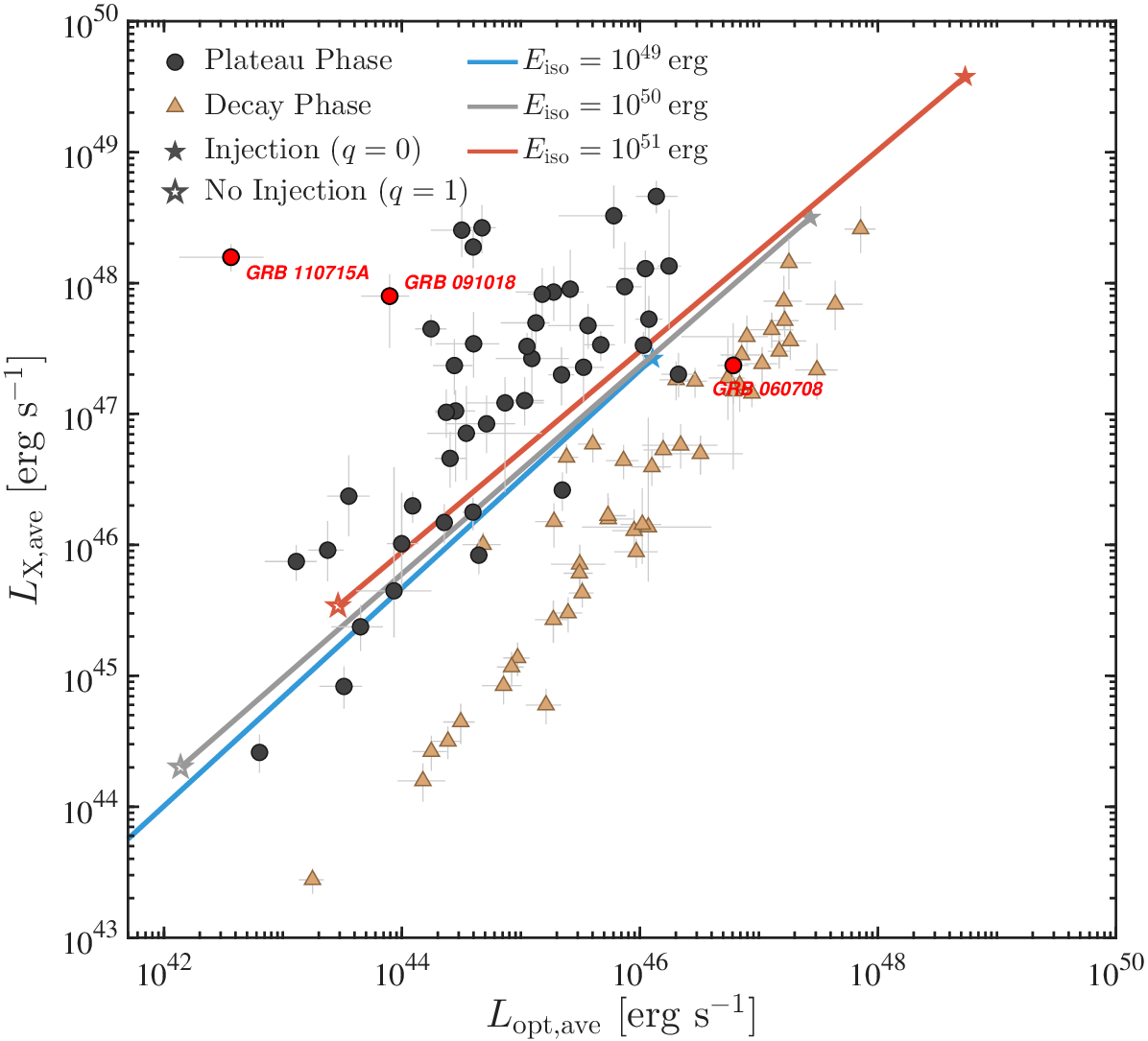}
\includegraphics[width=0.3\textwidth]{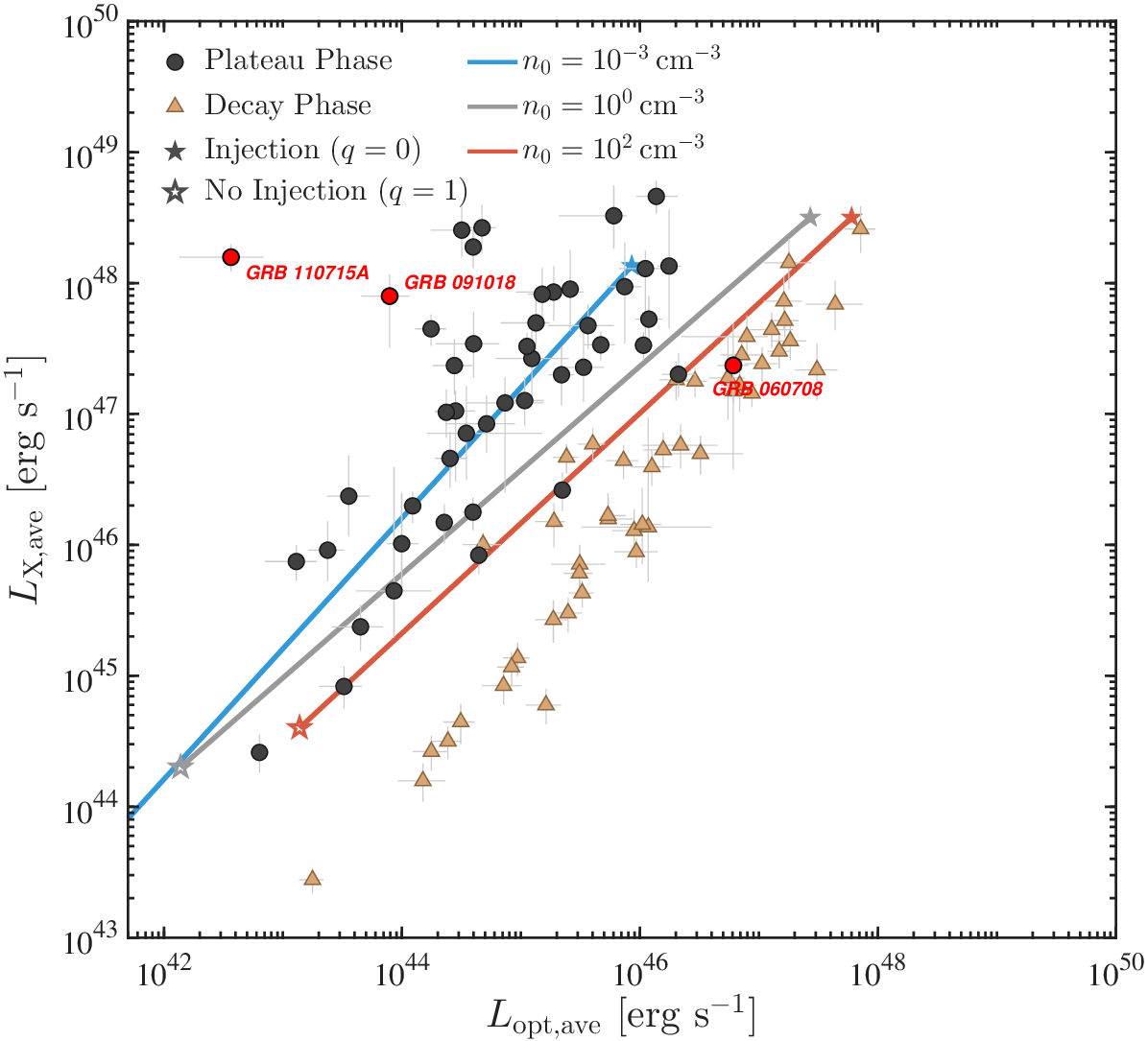}
\includegraphics[width=0.3\textwidth]{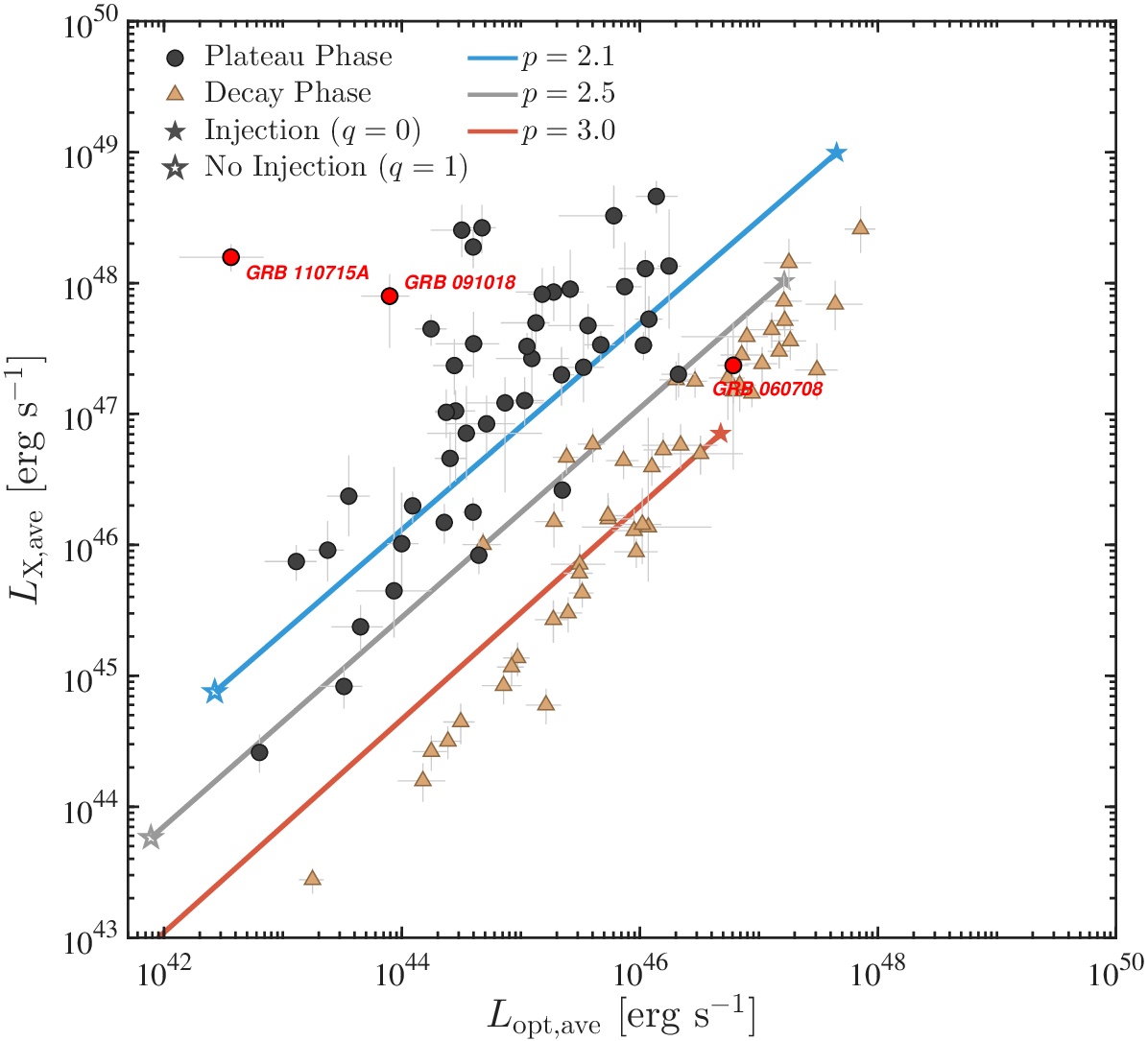}
\includegraphics[width=0.3\textwidth]{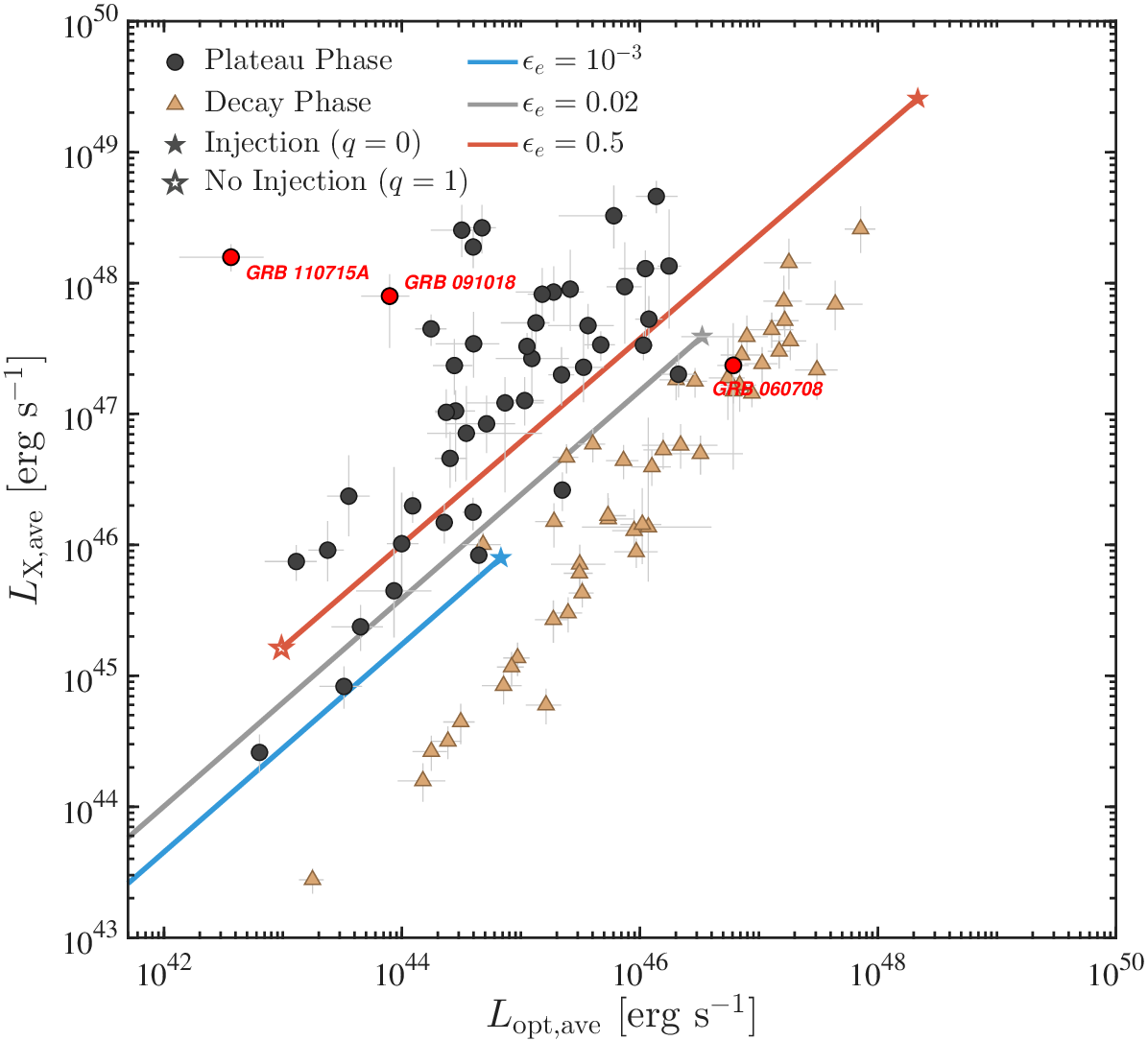}
\includegraphics[width=0.3\textwidth]{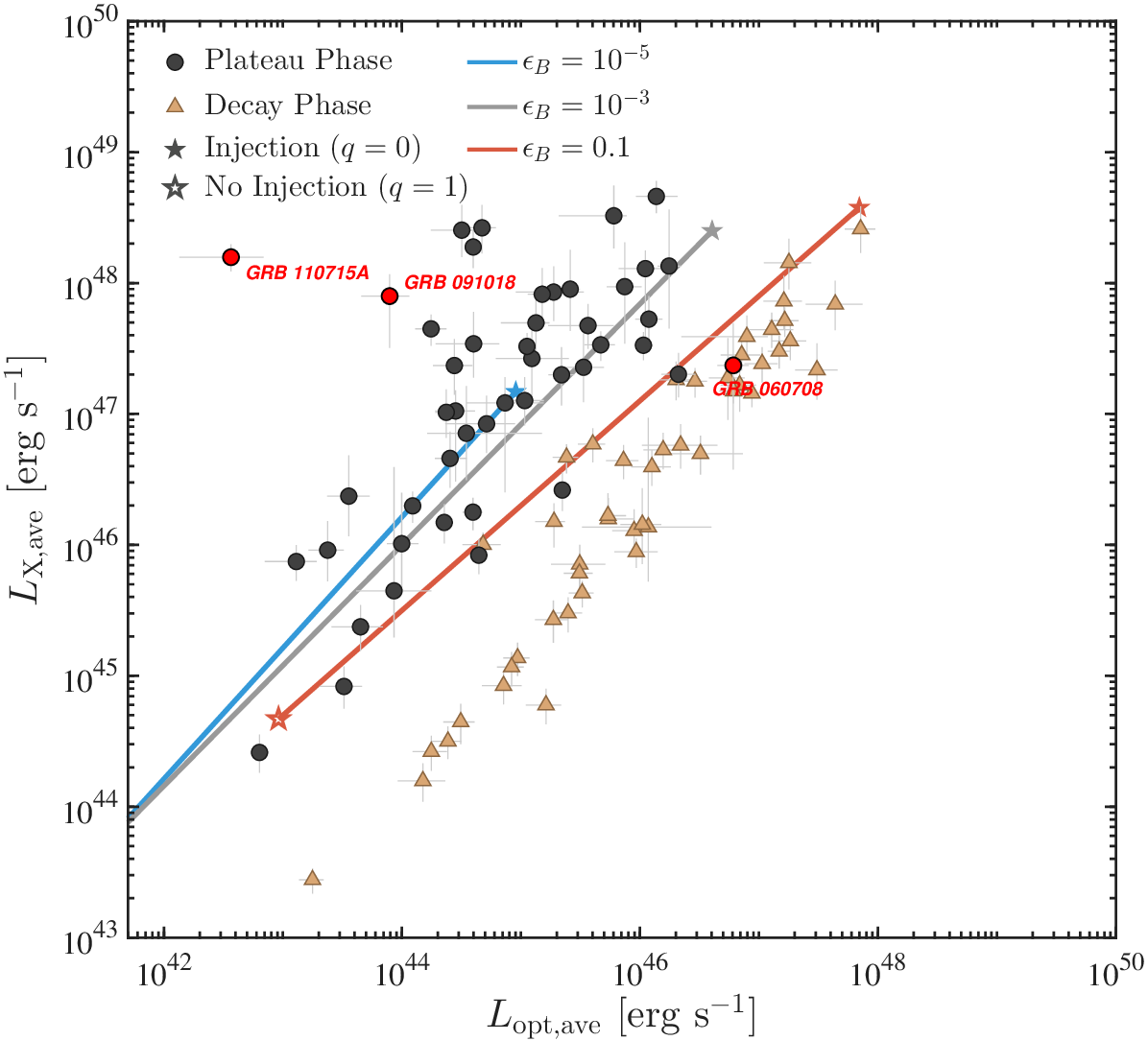}
\caption{Parameter space exploration of the energy injection model in $L_{\rm X, ave}$--$L_{\rm opt, ave}$ plane. Each panel demonstrates the effect of varying a single physical parameter, i.e., redshift $z$, isotropic kinetic energy $E_{\rm iso}$, ambient medium density $n_0$, electron spectral index $p$, and microphysical shock parameters $\epsilon_e$ and $\epsilon_B$. The other parameters are fixed as  $E_{\rm iso} = 10^{50} \, \rm erg$, $n_0 = 1.0~{\rm cm}^{-3}$, $\epsilon_e = 0.1$, $\epsilon_B = 0.01$, $p = 2.3$, and $z = 1.0$. The solid and open pentagrams represent the theoretical average luminosities for $q=0$ and $q=1$, respectively. The red dots represent the chromatic plateaus.
\label{MyFig2}}
\end{figure*}

\subsection{Parameter Space Exploration of the Energy Injection Model}
To investigate the physical origins of the observed $L_{\rm X, ave}$-- $L_{\rm opt, ave}$ correlation and the distribution of the plateau phase, we examined the sensitivity of the radiated luminosity to variations in key physical parameters. We calculated the theoretical average luminosities using the analytical afterglow model during the plateau phase, which is adopted from $100\,\mathrm{s}$ to $10^{5}\,\mathrm{s}$. 

For X-ray band, the instantaneous flux $F_{\rm X}$  is obtained by integrating the spectral energy distribution $F_{\nu}$ over the energy range of $0.3$--$10$~keV. For the optical band, we consider the R-band centered at frequency $\nu_{\rm R} = 4.56 \times 10^{14}$~Hz. The optical flux is calculated as $F_{\rm opt}(t) = F_{\nu}(\nu_{\rm R}, t) \Delta\nu_{\rm R}$, where $\Delta\nu_{\rm R} = 1.15 \times 10^{14}$~Hz is the effective bandwidth of the R-band filter.

In Figure~\ref{MyFig2}, the theoretical results are illustrated with pentagrams: the solid pentagrams represent the scenario with continuous energy injection (taking $q=0$), while the open pentagrams denote the case without energy injection ($q=1$). The connecting lines between the solid and open pentagrams indicate that the realistic injection processes in GRBs likely reside between these two theoretical limits. The sensitivity analysis was performed by varying one parameter at a time while keeping others at their fiducial values: redshift $z=1$, isotropic energy $E_{\rm iso} = 10^{50}\,\mathrm{erg}$, circumburst medium density $n_0 = 1\,\mathrm{cm^{-3}}$, microphysical energy equipartition fractions $\epsilon_e=0.1$ and $\epsilon_{\rm B}=0.1$, and the electron spectral index $p=2.3$.

In the top panels of Figure~\ref{MyFig2}, we show the relationships of calculated average luminosities with redshift $z$, isotropic energy $E_{\rm iso}$, and ambient density $n_0$. In the top-left panel, the luminosity increases with redshift $z$. One can find that the average luminosity is independent of $z$ for the energy injection case, while it is in proportion to the redshift without the energy injection, which can be derived directly from Equations~(\ref{eq:num}) - (\ref{Fnu_case}). In the top-middle panel, increasing the isotropic energy $E_{\rm iso}$ from $10^{49}$ to $10^{51}$~erg significantly shifts the theoretical average luminosity upward. This occurs because $E_{\rm iso}$ is a global scaling factor that positively affects both the peak frequency ($\nu_m \propto E_{\rm iso}^{1/2}$) and the peak flux density ($F_{\nu,\max} \propto E_{\rm iso}$), effectively lifting the entire synchrotron spectrum across all bands. Note that this $10^{49}$--$10^{51}$~erg range in the single-parameter analysis is selected to represent the typical energy scales of the bulk GRB afterglow population, where the late-time injected energy is generally smaller than or comparable to the initial kinetic energy of the burst itself. Plotting up to $10^{54}$~erg here would compress the typical data points in the plot, making the sensitivity tracks difficult to resolve. In contrast, in the global Monte Carlo simulation in Section~\ref{sec:results} and Figure~\ref{MyFig3}, we allow $E_{\rm iso}$ to vary up to $10^{54}$~erg to probe the physical upper boundary of the hyperaccreting BH engine. In the top-right panel, the effect of environment density $n_0$ is much more complex. As $n_0$ increases, the average luminosities at optical bands become higher ($L_{\rm opt, ave} \propto n_0^{1/2}$), while the X-ray luminosity remains relatively insensitive to density variations ($L_{\rm X, ave} \propto n_0^0$). Consequently, a denser environment pushes the emission tracks horizontally to the right in the $L_{\rm opt, ave}$--$L_{\rm X, ave}$ plane.

The bottom panels highlight the sensitivity to shock microphysics. The bottom-left panel shows the effect of the electron spectral index $p$. A harder spectrum (smaller $p$) retains more energy at high frequencies, leading to an increase in $L_{\rm X, ave}$ relative to $L_{\rm opt, ave}$. In the bottom-middle panel, higher $\epsilon_e$ values result in increased luminosity, which can be understood from the fact that $\epsilon_e$ governs the injection frequency ($\nu_m \propto \epsilon_e^2$); since the observing bands typically lie above $\nu_m$, the enhanced $\nu_m$ pushes more power into the radiative regime. Finally, as shown in the bottom-right panel, while $\epsilon_B$ positively correlates with the peak flux ($F_{\nu,\max} \propto \epsilon_B^{1/2}$), it also significantly reduces the cooling frequency ($\nu_c \propto \epsilon_B^{-3/2}$), leading to complex spectral shifts. 
The observational data for both the plateau phase and the decay phase are overplotted for comparison. However, varying a single parameter alone is insufficient to fully cover the entire scatter of the data points. To successfully reproduce the broad observational parameter space occupied by both plateau and decay phases, a simultaneous variation of multiple parameters is required. By allowing these parameters to vary within physically reasonable ranges, the energy injection model can successfully reproduce the entire observational parameter space occupied by both plateau and decay phases. 

\subsection{Observational Test for the Energy Injection Model}
To define the expected theoretical region for the plateau phase, we perform an enhanced Monte Carlo simulation with $N=2 \times 10^{5}$ random samples, ensuring a dense and robust coverage of the multidimensional parameter space. We adopt an energy injection index of $q=0$ and allow the physical parameters to vary within their typical broad ranges: the redshift $z \in [0.5, 5.0]$, the isotropic kinetic energy $E_{\rm iso} \in [10^{49}, 10^{54}] \rm\,erg$ (i.e., $E_{52} \in [10^{-3}, 100]$), the circumburst medium density $n_0 \in [0.01, 10] \rm\,cm^{-3}$, the microphysical parameters $\epsilon_e \in [10^{-3}, 0.5]$ and $\epsilon_B \in [10^{-5}, 0.01]$, and the electron spectral index $p \in[2.05, 3.0]$. 

\begin{figure}
\centering
\includegraphics[width=0.4\textwidth]{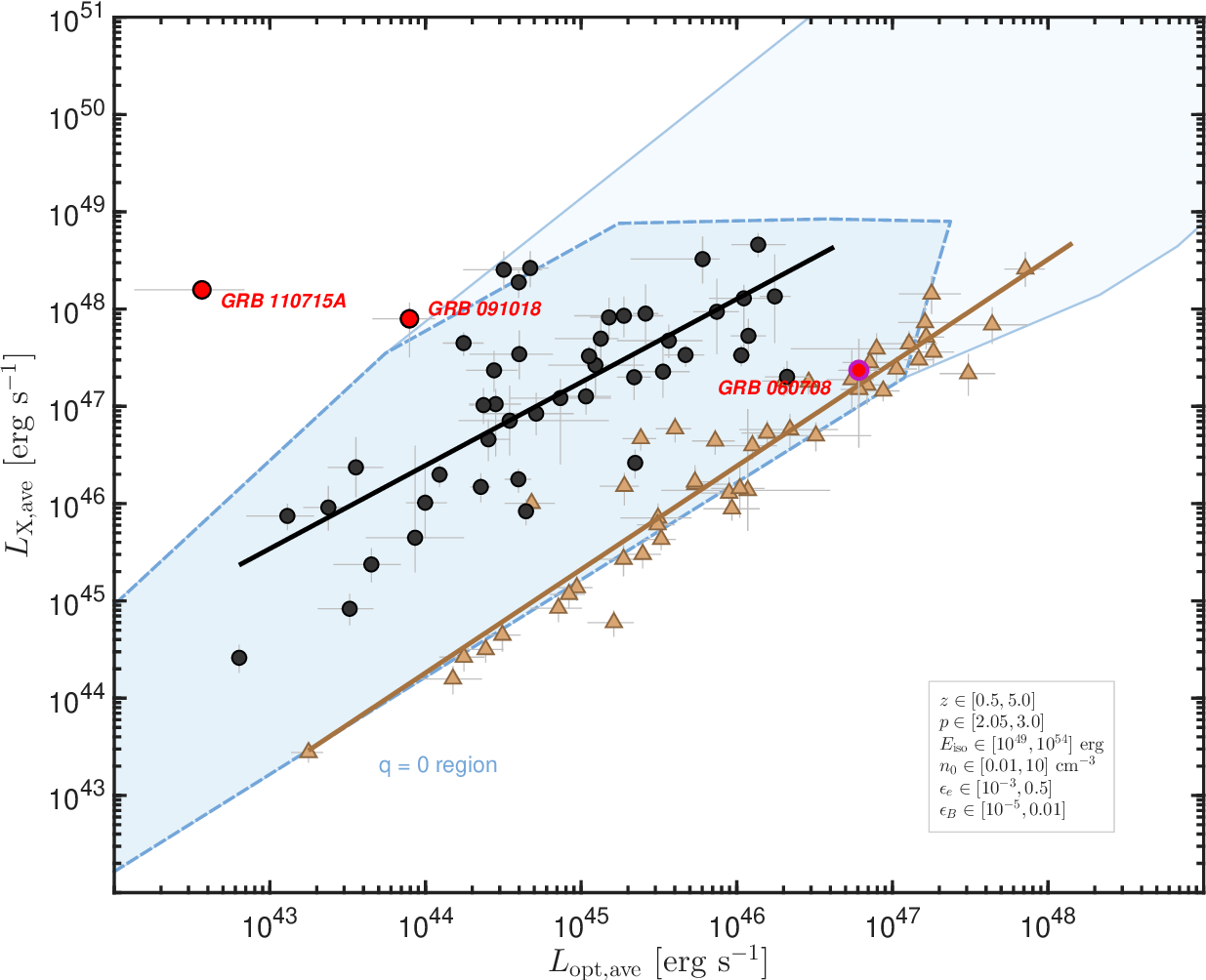}
\caption{Correlation between the isotropic average luminosities in X-ray and optical bands. The solid and dashed lines represent the expected theoretical region boundaries for a hyperaccreting BH engine limit ($E_{\rm iso} \le 10^{54}$~erg) and a magnetar engine limit ($E_{\rm iso} \le 10^{52}$~erg), respectively. The red dots represent the chromatic plateaus.}
\label{MyFig3}
\end{figure}

Crucially, we calculate the averaged luminosities to maintain consistency with the observational data processing. For the $q=0$ continuous energy injection scenario, the luminosity is integrated over the duration of the plateau phase, from $t_{\rm start} = 100$~s to the injection break time $T_{b}$. To account for the diversity of observed GRBs, $T_{b}$ is stochastically sampled within the range of $[10^2, 3 \times 10^5]$~s. The resulting ensemble of theoretical coordinates forms a dense distribution in the $L_{\rm opt, ave}$--$L_{\rm X, ave}$ plane. We then extract the outermost envelope of this distribution using a boundary-shrinking algorithm in the logarithmic space. In Figure~\ref{MyFig3}, we plot two theoretical boundaries: a solid line representing the BH engine limit ($E_{\rm iso} \le 10^{54}$~erg) and a dashed line representing the magnetar engine limit ($E_{\rm iso} \le 10^{52}$~erg). These regions represent the theoretically allowed parameter spaces for the continuous energy injection model under different central engines.

As shown in Figure~\ref{MyFig3}, the observational data for the plateau phase are largely confined within these theoretical boundaries. This consistency supports the interpretation that the plateau phase is powered by continuous energy injection from a long-lived central engine, although other scenarios cannot be completely ruled out. On the other hand, the chromatic plateaus (highlighted as red outliers) are situated at the very fringes or even outside of the theoretical region, which can hardly be explained by the energy injection model. Instead, these chromatic cases likely point toward more complex physical processes. Consequently, this $L_{\rm X, ave}$--$L_{\rm opt, ave}$ diagram provides a novel diagnostic method for certifying energy-injection-powered plateaus: sources that fall within the predicted $q=0$ region and follow the correlation relationship are candidates for the energy injection model.

Regarding the normal decay phase, the data points systematically shift downward compared with the points in the plateau phase while maintaining the same slope. This behavior is physically consistent with the cessation of energy injection: once the engine fades, the total energy in the blast wave stops increasing, causing a proportional drop in luminosity of all bands.

\section{Discussion and Conclusions}
\label{sec:summary} 
In this work, we utilized a comprehensive sample of 47 GRBs with simultaneous plateau detections in both X-ray and optical bands to probe the nature of the plateaus. By analyzing the average luminosity correlations of $L_{\rm X, ave}$ and $L_{\rm opt, ave}$ and comparing the observational data with the synchrotron forward shock model, we tested the validity of the energy injection model. Our main conclusions are summarized as follows.

We identified a moderate positive correlation between X-ray and optical average luminosities across both the plateau and normal decay phases. One can find that the slope of the correlation during the plateau phase ($m = 0.86 \pm 0.11$) is closely aligned with that of the subsequent normal decay phase ($m = 1.05 \pm 0.05$). This alignment provides compelling evidence that the plateau phase shares the same physical origin as the standard afterglow, i.e., synchrotron emission from the external forward shock, despite the difference in temporal decay indices.

By overplotting the theoretical parameter space, we demonstrated that the plateau data are largely confined within the boundaries predicted for a constant energy injection scenario ($q=0$).  This consistency is highly consistent with the scenario that the plateau phase is powered by continuous energy injection from the central engine, although other models cannot be completely ruled out. Consequently, the $L_{\rm X, ave}$--$L_{\rm opt, ave}$ diagram serves as a novel method to verify that plateaus that follow this correlation track and fall within the predicted region are candidates for the energy injection model. Conversely, relying on the temporal slope or closure relations alone may be insufficient \citep[as standard parameters like $p$, $q$, and the circumburst profile remain highly degenerate; see e.g.,][]{Racusin2009}. The plateaus that satisfy these criteria but lie outside the region shown in Figure~\ref{MyFig3} may not be powered by energy injection.

Furthermore, a clear evolutionary trend is observed regarding the transition from the plateau to the normal decay phase. The normal decay data exhibit a systematic downward shift in luminosity compared to the plateau phase, while retaining the same correlation slope. This reduction in luminosity is physically consistent with the central engine transitioning from a steady injection phase ($q=0$) to a cessation phase ($q \ge 1$). The cessation of the central engine reduces the total energy budget, leading to a downward shift in the normalization of the $L_{\rm X, ave}$--$L_{\rm opt, ave}$ correlation, while the underlying physical mechanism of the shock remains the same.

We note that the multi-wavelength plateau luminosity correlation has also been studied in other works. For instance, \citet{Lenart2025} extended it into a three-dimensional (3D) correlation (a fundamental plane) by introducing the rest-frame plateau duration $T_a$ as the third parameter. This 3D relation breaks parameter degeneracies and tests energy conservation for the central engine (e.g., constraining the magnetar rotational energy $E_{\rm rot} \lesssim 2 \times 10^{52}$~erg or BH accretion/spin-down efficiency; \citealt{Kumar2008, Cannizzo2009, Cannizzo2011}). Compared with these relations focusing on the break point, our work uses the time-averaged luminosity $L_{\rm ave}$ integrated over each phase. This enables a self-consistent comparison between the plateau and normal decay phases. By showing that the correlation slope remains nearly invariant, we verify the shared synchrotron shock origin of both phases, demonstrating physical continuity across the temporal break.

Furthermore, \citet{vanEerten2014a} and \citet{vanEerten2014b} modeled the impact of energy injection on the luminosity-duration correlation, where the plateau duration ($T_b$) represents the active energy supply phase linked to the total energy budget ($E_{\rm inj} \propto L  T_b$). In this framework, the achromatic break times in our sample ($T_{\rm b, X} \approx T_{\rm b, o}$) favor a central engine origin over spectral transitions, which would otherwise produce chromatic breaks \citep{vanEerten2014a}. Combining our luminosity relations with the duration distribution can thus help constrain the lifetime and structure of the central engine.

We can also compare our findings with our previous work \citep{2026ApJ...998..298L}, which tested the energy-injection model using standard closure relations on the same sample. In \citet{2026ApJ...998..298L}, the forward shock origin was supported by analyzing individual temporal and spectral indices. The current work complements that study by introducing a global perspective based on time-averaged luminosity correlations and Monte Carlo parameter envelopes, which avoids the degeneracies of individual closure relation tests. In addition, showing that the correlation slope remains nearly invariant across the break ($0.86$ versus $0.97$) verifies the physical continuity and shared forward shock origin of both phases. Together, these two complementary methodologies provide a more robust verification of the energy-injection model.

In summary, our results demonstrate that multi-band observations are critical for identifying the nature of the plateau. The synchronized evolution of optical and X-ray average luminosities serves as a diagnostic to distinguish the energy injection model using the achromatic breaks. This study suggests that the plateau phase corresponds to a period of active energy supply from the central engine, while the normal decay phase follows naturally once this injection ceases.

It is crucial to note the role of parameter degeneracy when interpreting the $L_{\rm X, ave}$--$L_{\rm opt, ave}$ diagram. Due to the high dimensionality of the standard afterglow parameter space, a ``chromatic'' single-band plateau (e.g., a plateau in X-rays but a normal decay in the optical band) might coincidentally fall into the $q=0$ theoretical region.

However, this degeneracy may be effectively broken by tracing the temporal evolutionary trajectory of the individual burst within the $L_{\rm X}(t)$--$L_{\rm opt}(t)$ plane. For a genuine achromatic energy injection (multi-band plateau), the evolutionary track of the burst over time will evolve with a slope parallel to the ensemble correlation ($m \approx 0.8$). Conversely, a chromatic single-band plateau may show a different track, for instance, moving horizontally to the left if only the X-ray band is in a plateau phase while the optical band decays rapidly. Further progress in this direction will benefit from a larger sample of high-quality, simultaneous multi-band observations.

\section*{acknowledgments}
We thank Chen Deng for helpful discussion. This work was supported by the National Natural Science Foundation of China (Grant Nos. 12494572, 12221003, and 12503052).


\begin{thebibliography}{}
\bibitem[Burrows et al.(2005)]{2005SSRv..120..165B} Burrows, D.~N., Hill, J.~E., Nousek, J.~A., et al.\ 2005, \ssr, 120, 3-4, 165. doi:10.1007/s11214-005-5097-2
\bibitem[Cannizzo \& Gehrels(2009)]{Cannizzo2009} Cannizzo, J.~K., \& Gehrels, N.\ 2009, \apj, 700, 1047. doi:10.1088/0004-637X/700/2/1047
\bibitem[Cannizzo et al.(2011)]{Cannizzo2011} Cannizzo, J.~K., Troja, E., \& Gehrels, N.\ 2011, \apj, 734, 35. doi:10.1088/0004-637X/734/1/35
\bibitem[Dai \& Lu(1998)]{1998AA...333L..87D} Dai, Z.~G. \& Lu, T.\ 1998, \aap, 333, L87. doi:10.48550/arXiv.astro-ph/9810402
\bibitem[Dainotti et al.(2016)]{2016ApJ...825L..20D} Dainotti, M.~G., Postnikov, S., Hernandez, X., et al.\ 2016, \apjl, 825, 2, L20. doi:10.3847/2041-8205/825/2/L20
\bibitem[Dainotti et al.(2020)]{2020ApJ...904...97D} Dainotti, M.~G., Lenart, A. {\L}., Sarracino, G., et al.\ 2020, \apj, 904, 2, 97. doi:10.3847/1538-4357/abbe8a
\bibitem[Deng et al.(2026)]{2026ApJ..1000...97D} Deng, C., Huang, Y.-F., Kurban, A., et al.\ 2026, \apj, 1000, 1, 97. doi:10.3847/1538-4357/ae486b
\bibitem[de Wet et al.(2023)]{2023AA...671A.116D} de Wet, S., Laskar, T., Groot, P.~J., et al.\ 2023, \aap, 671, A116. doi:10.1051/0004-6361/202244917
\bibitem[Dereli-B{\'e}gu{\'e} et al.(2022)]{2022NatCo..13.5611D} Dereli-B{\'e}gu{\'e}, H., Pe'er, A., Ryde, F., et al.\ 2022, Nature Communications, 13, 5611. doi:10.1038/s41467-022-32881-1
\bibitem[Duffell \& MacFadyen(2015)]{2015ApJ...806..205D} Duffell, P.~C. \& MacFadyen, A.~I.\ 2015, \apj, 806, 2, 205. doi:10.1088/0004-637X/806/2/205
\bibitem[Eichler \& Granot(2006)]{2006ApJ...641L...5E} Eichler, D. \& Granot, J.\ 2006, \apjl, 641, 1, L5. doi:10.1086/503667
\bibitem[Fan \& Piran(2006)]{2006MNRAS.369..197F} Fan, Y. \& Piran, T.\ 2006, \mnras, 369, 1, 197. doi:10.1111/j.1365-2966.2006.10280.x
\bibitem[Fan et al.(2013)]{2013ApJ...779L..25F} Fan, Y.-Z., Yu, Y.-W., Xu, D., et al.\ 2013, \apjl, 779, 2, L25. doi:10.1088/2041-8205/779/2/L25
\bibitem[Gao et al.(2013)]{2013NewAR..57..141G} Gao, H., Lei, W.-H., Zou, Y.-C., et al.\ 2013, \nar, 57, 6, 141. doi:10.1016/j.newar.2013.10.001
\bibitem[Gehrels et al.(2004)]{2004ApJ...611.1005G} Gehrels, N., Chincarini, G., Giommi, P., et al.\ 2004, \apj, 611, 2, 1005. doi:10.1086/422091
\bibitem[Granot et al.(2006)]{2006MNRAS.370.1946G} Granot, J., K{\"o}nigl, A., \& Piran, T.\ 2006, \mnras, 370, 4, 1946. doi:10.1111/j.1365-2966.2006.10621.x
\bibitem[Hou et al.(2018)]{2018ApJ...854..104H} Hou, S.-J., Liu, T., Xu, R.-X., et al.\ 2018, \apj, 854, 2, 104. doi:10.3847/1538-4357/aaabba
\bibitem[Huang \& Liu(2021)]{2021ApJ...916...71H} Huang, B.-Q. \& Liu, T.\ 2021, \apj, 916, 2, 71. doi:10.3847/1538-4357/ac07a0
\bibitem[Huang \& Liu(2024)]{2024Univ...10..438H} Huang, B.-Q. \& Liu, T.\ 2024, Universe, 10, 12, 438. doi:10.3390/universe10120438
\bibitem[Huang et al.(2012)]{2012grb..confE..75H} Huang, K., Chiu, K.-L., \& Urata, Y.\ 2012, Proceedings of Science, Gamma-Ray Bursts 2012 Conference (GRB 2012), 152, 75. doi:10.22323/1.152.0075
\bibitem[Ioka et al.(2006)]{2006AA...458....7I} Ioka, K., Toma, K., Yamazaki, R., et al.\ 2006, \aap, 458, 1, 7. doi:10.1051/0004-6361:20064939
\bibitem[Jin et al.(2007)]{2007ApJ...656L..57J} Jin, Z.~P., Yan, T., Fan, Y.~Z., et al.\ 2007, \apjl, 656, 2, L57. doi:10.1086/512971
\bibitem[Kagawa et al.(2019)]{2019ApJ...877..147K} Kagawa, Y., Yonetoku, D., Sawano, T., et al.\ 2019, \apj, 877, 2, 147. doi:10.3847/1538-4357/ab1bd6
\bibitem[Kobayashi \& Zhang(2007)]{2007ApJ...655..973K} Kobayashi, S. \& Zhang, B.\ 2007, \apj, 655, 2, 973. doi:10.1086/510203
\bibitem[Kumar et al.(2008)]{Kumar2008} Kumar, P., Narayan, R., \& Johnson, J.~L.\ 2008, \mnras, 388, 1729. doi:10.1111/j.1365-2966.2008.13524.x
\bibitem[Lenart et al.(2025)]{Lenart2025} Lenart, A.~{\L}., Dainotti, M.~G., Khatiya, N.~S., et al.\ 2025, Journal of High Energy Astrophysics, 47, 100384. doi:10.1016/j.jheap.2025.100384
\bibitem[Li et al.(2016)]{2016PhRvD..94h3010L} Li, A., Zhang, B., Zhang, N.-B., et al.\ 2016, \prd, 94, 8, 083010. doi:10.1103/PhysRevD.94.083010
\bibitem[Li et al.(2012)]{2012ApJ...758...27L} Li, L., Liang, E.-W., Tang, Q.-W., et al.\ 2012, \apj, 758, 1, 27. doi:10.1088/0004-637X/758/1/27
\bibitem[Li et al.(2015)]{2015ApJ...805...13L} Li, L., Wu, X.-F., Huang, Y.-F., et al.\ 2015, \apj, 805, 1, 13. doi:10.1088/0004-637X/805/1/13
\bibitem[Li et al.(2021)]{2021ApJ...922...22L} Li, X.-Y., Lin, D.-B., Ren, J., et al.\ 2021, \apj, 922, 1, 22. doi:10.3847/1538-4357/ac1ff2
\bibitem[Li et al.(2026)]{2026ApJ...998..298L} Li, X.-Y., Liu, T., Huang, B.-Q., et al.\ 2026, \apj, 998, 2, 298. doi:10.3847/1538-4357/ae346b
\bibitem[Liang et al.(2007)]{2007ApJ...670..565L} Liang, E.-W., Zhang, B.-B., \& Zhang, B.\ 2007, \apj, 670, 1, 565. doi:10.1086/521870
\bibitem[Liu et al.(2017)]{2017NewAR..79....1L} Liu, T., Gu, W.-M., \& Zhang, B.\ 2017, \nar, 79, 1. doi:10.1016/j.newar.2017.07.001
\bibitem[L{\"u} \& Zhang(2014)]{2014ApJ...785...74L} L{\"u}, H.-J. \& Zhang, B.\ 2014, \apj, 785, 1, 74. doi:10.1088/0004-637X/785/1/74
\bibitem[M{\'e}sz{\'a}ros \& Rees(1997)]{1997ApJ...476..232M} M{\'e}sz{\'a}ros, P. \& Rees, M.~J.\ 1997, \apj, 476, 1, 232. doi:10.1086/303625
\bibitem[Nappo et al.(2017)]{2017AA...598A..23N} Nappo, F., Pescalli, A., Oganesyan, G., et al.\ 2017, \aap, 598, A23. doi:10.1051/0004-6361/201628801
\bibitem[Nousek et al.(2006)]{2006ApJ...642..389N} Nousek, J.~A., Kouveliotou, C., Grupe, D., et al.\ 2006, \apj, 642, 1, 389. doi:10.1086/500724
\bibitem[Oates et al.(2011)]{Oates2011} Oates, S.~R., Page, M.~J., Schady, P., et al.\ 2011, \mnras, 412, 561. doi:10.1111/j.1365-2966.2010.17926.x
\bibitem[O'Brien et al.(2006)]{2006ApJ...647.1213O} O'Brien, P.~T., Willingale, R., Osborne, J., et al.\ 2006, \apj, 647, 2, 1213. doi:10.1086/505457
\bibitem[Panaitescu et al.(2006a)]{2006MNRAS.369.2059P} Panaitescu, A., M{\'e}sz{\'a}ros, P., Burrows, D., et al.\ 2006a, \mnras, 369, 4, 2059. doi:10.1111/j.1365-2966.2006.10453.x
\bibitem[Panaitescu et al.(2006b)]{2006MNRAS.366.1357P} Panaitescu, A., M{\'e}sz{\'a}ros, P., Gehrels, N., et al.\ 2006b, \mnras, 366, 4, 1357. doi:10.1111/j.1365-2966.2005.09900.x
\bibitem[Racusin et al.(2009)]{Racusin2009} Racusin, J.~L., Liang, E.~W., Zhang, B., et al.\ 2009, \apj, 698, 43. doi:10.1088/0004-637X/698/1/43
\bibitem[Ronchini et al.(2023)]{2023AA...675A.117R} Ronchini, S., Stratta, G., Rossi, A., et al.\ 2023, \aap, 675, A117. doi:10.1051/0004-6361/202245348
\bibitem[Sari et al.(1998)]{1998ApJ...497L..17S} Sari, R., Piran, T., \& Narayan, R.\ 1998, \apjl, 497, 1, L17. doi:10.1086/311269
\bibitem[Toma et al.(2006)]{2006ApJ...640L.139T} Toma, K., Ioka, K., Yamazaki, R., et al.\ 2006, \apjl, 640, 2, L139. doi:10.1086/503384
\bibitem[van Eerten(2014a)]{vanEerten2014a} van Eerten, H.\ 2014a, \mnras, 442, 3495. doi:10.1093/mnras/stu1025
\bibitem[van Eerten(2014b)]{vanEerten2014b} van Eerten, H.\ 2014b, \mnras, 445, 2414. doi:10.1093/mnras/stu1921
\bibitem[Wang et al.(2015)]{2015ApJS..219....9W} Wang, X.-G., Zhang, B., Liang, E.-W., et al.\ 2015, \apjs, 219, 1, 9. doi:10.1088/0067-0049/219/1/9
\bibitem[Yamazaki(2009)]{2009ApJ...690L.118Y} Yamazaki, R.\ 2009, \apjl, 690, 2, L118. doi:10.1088/0004-637X/690/2/L118
\bibitem[Yi et al.(2022)]{2022ApJ...924...69Y} Yi, S.-X., Du, M., \& Liu, T.\ 2022, \apj, 924, 2, 69. doi:10.3847/1538-4357/ac35e7
\bibitem[Zhang \& M{\'e}sz{\'a}ros(2001)]{2001ApJ...552L..35Z} Zhang, B. \& M{\'e}sz{\'a}ros, P.\ 2001, \apjl, 552, 1, L35. doi:10.1086/320255
\bibitem[Zhang \& M{\'e}sz{\'a}ros(2004)]{Zhang2004} Zhang, B. \& M{\'e}sz{\'a}ros, P.\ 2004, International Journal of Modern Physics A, 19, 2380. doi:10.1142/S0217751X0401746X
\bibitem[Zhang et al.(2006)]{2006ApJ...642..354Z} Zhang, B., Fan, Y.~Z., Dyks, J., et al.\ 2006, \apj, 642, 1, 354. doi:10.1086/500723
\bibitem[Zhang et al.(2007)]{2007ApJ...666.1002Z} Zhang, B.-B., Liang, E.-W., \& Zhang, B.\ 2007, \apj, 666, 2, 1002. doi:10.1086/519548
\bibitem[Zhu et al.(2023)]{2023ApJ...948...30Z} Zhu, Z.-P., Xu, D., Fynbo, J.~P.~U., et al.\ 2023, \apj, 948, 1, 30. doi:10.3847/1538-4357/acbd96
\end{thebibliography}
\end{document}